\documentclass[10pt,journal]{IEEEtran}
\usepackage{amsmath,amssymb,amsfonts}
\usepackage{algorithmic}
\usepackage{algorithm}
\usepackage{array}
\usepackage[caption=false,font=footnotesize,labelfont=sf,textfont=sf]{subfig}
\usepackage{textcomp}
\usepackage{stfloats}
\usepackage{url}
\usepackage{verbatim}
\usepackage{graphicx}
\usepackage{cite}

\usepackage{bm}
\usepackage{bbm}
\usepackage{amsthm}

\newtheorem{proposition}{Proposition}

\newtheorem{definition}{Definition}

\graphicspath{{figs/}}

\newcommand{\nv}{\mathbf{n}}

\newcommand{\pv}{\mathbf{p}}

\newcommand{\uv}{\mathbf{u}}
\newcommand{\wv}{\mathbf{w}}

\newcommand{\xv}{\mathbf{x}}
\newcommand{\yv}{\mathbf{y}}

\newcommand{\Bm}{\mathbf{B}}
\newcommand{\Cm}{\mathbf{C}}
\newcommand{\Dm}{\mathbf{D}}

\newcommand{\Id}{\mathbf{I}}

\newcommand{\Lm}{\mathbf{L}}

\newcommand{\Tm}{\mathbf{T}}
\newcommand{\Um}{\mathbf{U}}
\newcommand{\Wm}{\mathbf{W}}
\newcommand{\Vm}{\mathbf{V}}

\newcommand{\Ec}{\mathcal{E}}
\newcommand{\Fc}{\mathcal{F}}
\newcommand{\Gc}{\mathcal{G}}

\newcommand{\Nc}{\mathcal{N}}
\newcommand{\Oc}{\mathcal{O}}

\newcommand{\Sc}{\mathcal{S}}

\newcommand{\Vc}{\mathcal{V}}

\newcommand{\Lambdam}{\hbox{\boldmath$\Lambda$}}

\begin{document}

\title{Edge Sampling of Graphs: Graph Signal Processing Approach With Edge Smoothness}

\author{Kenta Yanagiya,~\IEEEmembership{Graduate Student Member,~IEEE,} Koki Yamada,~\IEEEmembership{Member,~IEEE,} Yasuo Katsuhara,\\Tomoya Takatani, and Yuichi Tanaka,~\IEEEmembership{Senior Member,~IEEE}
\thanks{The preliminary version of this paper has been presented in \cite{yanagiya2022}.}
\thanks{Kenta Yanagiya and Yuichi Tanaka are with the Graduate School of Engineering, Osaka University, Suita, Osaka 565-0861, Japan (e-mail: k\_yanagiya@msp-lab.org; ytanaka@comm.eng.osaka-u.ac.jp). Koki Yamada is with the Graduate School of Electrical Engineering and Computer Science, Tokyo University of Agriculture and Technology, Koganei, Tokyo 184-8588, Japan (e-mail: k-yamada@go.tuat.ac.jp). Yasuo Katsuhara and Tomoya Takatani are with the Toyota Motor Corporation, Susono, Shizuoka 410-1107, Japan (e-mail: yasuo\_katsuhara@mail.toyota.co.jp; tomoya\_takatani@konpon.toyota).}}

\markboth{Journal of \LaTeX\ Class Files,~Vol.~14, No.~8, August~2021}%
{Shell \MakeLowercase{\textit{et al.}}: A Sample Article Using IEEEtran.cls for IEEE Journals}

\maketitle

\begin{abstract}
Finding important edges in a graph is a crucial problem for various research fields, such as network epidemics, signal processing, machine learning, and sensor networks.
In this paper, we tackle the problem based on sampling theory on graphs.
We convert the original graph to a \textit{line graph} where its nodes and edges, respectively, represent the original edges and the connections between the edges.
We then perform node sampling of the line graph based on the \textit{edge smoothness} assumption: This process selects the most critical edges in the original graph.
We present a general framework of edge sampling based on graph sampling theory and reveal a theoretical relationship between the degree of the original graph and the line graph.
We also propose an acceleration method for edge sampling in the proposed framework by using the relationship between two types of Laplacian of the node and edge domains.
Experimental results in synthetic and real-world graphs validate the effectiveness of our approach against some alternative edge selection methods.
\end{abstract}


\section{Introduction}
\IEEEPARstart{G}{raph} signal processing (GSP) is a developing field of signal processing \cite{Shuman2013, Ortega2018, Cheung2018, Tanaka2020c}.
GSP targets \textit{graph signals} whose domain is represented as nodes of a graph.
There exist various promising applications of GSP since many real-world signals have underlying structures beyond the standard uniform-interval relationships.

Examples of graph signals include signals on social/brain/transportation/power networks and point clouds \cite{Tanaka2018, Tanaka2020b, Egilme2017, Yamada2020, sakiyama2019, Tanaka2014a, Sakiya2014a, Cheung2018, Onuki2016}.
GSP aims to extend theories and algorithms for standard signal processing like sampling, filtering, restoration, and compression.

Graph signal sampling is a counterpart of that for standard, i.e., uniform-interval, signals \cite{Tanaka2020c}.
Standard signals implicitly assume their structure because the sampling period is determined prior to sampling and is usually fixed throughout the sampling process.
In contrast, graph signals could have irregular structures, and their corresponding graph frequencies are unevenly distributed.
The optimal sampling of graph signals often depends on graphs, especially for noisy cases.
Therefore, sampling methods on graphs have been studied extensively \cite{Anis2016, Chen2015, Marque2016, sakiyama2019, Tanaka2020c, Tanaka2018, Tanaka2020b, Hara2020, tsitsvero2016}.
However, most existing sampling methods on GSP consider the sampling of \textit{nodes} as an analogy of that in standard signal processing.
There exists another entity for sampling in GSP: \textit{Edges}.

In this paper, we propose \textit{edge sampling} based on graph signal processing techniques.
As an analogy to node sampling, edge sampling refers to selecting the most essential edges from a given set of edges in the original graph.
A new graph comprises the original nodes and sampled, i.e., selected, edges.

Edge sampling is required in various application fields.
For example, in network epidemiology, we need to select the essential edges for preventing disease spreading by removing these edges \cite{nowzari2016,nishi2020}.
This technique is important for policymakers to determine an effective lockdown policy \cite{chang2021a}.
For network science, we often need to obtain a good abstraction of graphs, i.e., edge sparsification \cite{spielman2011,spielman2011spectral,fung2019}, to save computation and storage burden.

In this paper, we view edge sampling from a GSP perspective.
In the proposed approach, we assume the \textit{smoothness} of edge weights.
The smoothness of edge weights refers to the similarity of the weights of adjacent edges, i.e., edges connected to the same node.
As we will show later, this edge smoothness can be found in various graphs.
We utilize \textit{graph frequency} to measure the smoothness.
However, the smoothness in GSP usually refers to the \textit{smoothness of the signal} on the nodes.
To properly regard edge smoothness as signal smoothness, we convert the original graph into a \textit{line graph} that represents the edge connection relationship of the original graph.
We select nodes in the line graph based on a GSP technique.
As a result, selecting important nodes in the line graph is regarded as selecting important edges in the original graph.
We can use various effective and high-quality node sampling methods of GSP by converting to the line graph \cite{sakiyama2019, Anis2016, tsitsvero2016}.

We also propose an acceleration method of edge sampling.
Edge sampling often requires matrix multiplication(s)
to reconstruct edge weights by filtering similar to the signal sampling counterpart.
Since we treat edge weights as graph signals in the edge sampling, the size of the filter matrix is identical to the number of edges: It is generally larger than the number of nodes.
As a result, the most calculation-intensive part of the edge sampling process is filtering.
We propose an approximation method for the filtering process to avoid large matrix multiplication that does not require explicit line graph conversion.

In the experiments of synthetic and real-world graphs, our proposed edge sampling method outperforms alternative edge selection methods regarding several measures.

The remainder of this paper is organized as follows.
Section \ref{sec: related_work} reviews the sampling of graph signals and edge sparsification/reduction.
In Section \ref{sec: edge_sampling}, the framework of the proposed method is described along with smoothness prior in the edge domain.
An acceleration method for edge sampling is proposed in Section \ref{sec: acceleration}.
Section \ref{sec: experiments} presents numerical experiments on synthetic and real data.
Finally, conclusions are given in Section \ref{sec: conclusion}.

\noindent
\textit{Notation:}
A graph is denoted as $\Gc = (\Vc, \Ec)$, where $\Vc$ and $\Ec$ are the sets of nodes and edges, respectively. 
The number of nodes is $N=|\Vc|$ unless otherwise specified. 
The adjacency matrix of $\Gc$ is represented as $\mathbf{W}$, where its $(m,n)$-entry $w_{mn}$ is the edge weight between nodes $m$ and $n$; $w_{mn} = 0$ for unconnected nodes. 
The degree matrix $\Dm$ is diagonal, with $m$th diagonal element  $[\Dm]_{mm} = d_m := \sum_n w_{mn}$.
In this paper, we consider undirected graphs without self-loops, i.e., $[\Wm]_{nm} = [\Wm]_{mn}$ and $[\Wm]_{nn} = 0$ for all $m$ and $n$.
In addition to the (weighted) degree, the number of edges connecting to the node $m$, i.e., unweighted degree, is defined as $k_m:= \sum_n \mathbbm{1}_{(\text{if node } m \text{ is connected to node } n)}$.

For an unweighted graph, the incidence matrix $\Bm \in \mathbb{R}^{N \times |\Ec|}$ is defined as follows:
\begin{equation}
[\Bm]_{i\alpha} =  \begin{cases}
1 & \text{Edge } \alpha \textrm{ is incident to node } i,\\
0 & \text{Otherwise}.
\end{cases}
\label{eqn: incidence_unweighted}
\end{equation}
In other words, each column in $\Bm$ represents an edge, and two nonzero elements in each column correspond to the nodes connecting by the edge.
The incidence matrix for a weighted graph $\tilde{\Bm}$ is similarly defined with replacing $1$ in \eqref{eqn: incidence_unweighted} by $\sqrt{w_{\alpha}}$,
where $w_{\alpha}$ is the weight of the edge $\alpha$.
In this paper, in addition to the incidence matrix described above, we also define a directed incidence matrix as follows:
\begin{equation}
[\bar{\Bm}]_{i\alpha}  = \begin{cases}
\sqrt{w_{\alpha}} & \text{Edge } \alpha \textrm{ is incident to node }i,\\
-\sqrt{w_{\alpha}} & \text{Edge } \alpha \textrm{ is incident to node }j,\\
0 & \text{Otherwise}.
\end{cases}
\label{eqn: incidence_directed}
\end{equation}
Note that $\tilde{\Bm} = |\bar{\Bm}|$, where $|\cdot|$ makes each element of the given matrix into the absolute value.
Directed incidence matrices can also be defined for undirected graphs by considering pseudo-orientation\footnote{The pseudo-orientation is induced by the lexicographic order of incident nodes $i$ and $j$, i.e., edges are oriented from the node $i$ to the node $j$ ($i<j$) \cite{schaub2018}.}.

Graph Laplacian is defined as $\Lm:=\Dm - \Wm$.
Note that $\Lm = \bar{\Bm}\bar{\Bm}^\top$.
Since $\Lm$ is a real symmetric matrix, it always has an eigendecomposition $\Lm = \Um \bm{\Lambda} \Um^\top$, where $\Um = [\uv_0, \ldots, \uv_{N-1}]$ is an orthonormal matrix containing the eigenvectors $\uv_i$, and $\bm{\Lambda} = \text{diag}(\lambda_0, \ldots, \lambda_{N-1})$ consists of the eigenvalues $\lambda_i$. 
These eigenvalues are assumed to be ordered as $0=\lambda_0<\lambda_1 \leq \lambda_2 \cdots \leq \lambda_{N-1}=\lambda_{\max}$ without loss of generality.
We refer to $\lambda_i$ as the \textit{graph frequency}. 
A graph signal $x: \Vc \rightarrow \mathbb{R}$ is a function that assigns a real value to each node.
Graph signals can be written as vectors $\xv \in \mathbb{R}^N$ whose $n$th element, $[\xv]_n$, represents the signal value at the $n$th node. 
The graph Fourier transform (GFT) is defined as $[\hat{\xv}]_i = \langle \uv_{i}, \xv\rangle = \sum_{n=0}^{N-1}[\uv_{i}]_n[\xv]_n.$

\section{Related Work}\label{sec: related_work}
In this section, we briefly review existing approaches of sampling theory on graphs and edge sparsification/reduction.

\subsection{Graph Signal Sampling}
Sampling of graph signals is a fundamental topic on GSP \cite{Tanaka2020c, Chen2015, Anis2016}.
Node sampling selects samples on $\Sc \subset \Vc$ where $\Sc$ is called a \textit{sampling set}.
Since there is no ``regular sampling'' in general in the graph setting, the sampling quality depends on $\Sc$, especially in the noisy case.
Therefore, various sampling strategies have been proposed so far \cite{Tanaka2018, Hara2020, Tanaka2020b, sakiyama2019, Tanaka2020c}.

Note that there have been no edge sampling methods based on graph sampling theory so far because existing methods focus on reducing the number of samples based on the assumption of signal smoothness.

\subsection{Edge Sparsification and Reduction}
Removing edges in a graph has been studied in many fields, such as graph theory and network science.
There are various methods with different motivations.

In graph theory, the reduction of edges is often called \textit{edge sparsification} \cite{spielman2011,spielman2011spectral, fung2019}.
The motivation of edge sparsification is to preserve characteristics of the original graph, e.g., eigenvalue distribution and connectivity, in the modified graph.
Although there exist theoretical guarantees, edge sparsification methods have three major issues in real applications.
First, they often have to modify edge weights, i.e., the edge weights in the sparsified graph are changed.
Second, the number of removed edges cannot be determined before sparsification.
Even under the same parameter(s), the number of removed edges can vary due to randomness in the algorithms.
Third, when a random selection method is considered, the importance of selected edges is not ordered.
In other words, if further edges are added/removed in a manipulated graph,  the importance of previously selected edges is not available and the whole process of random selection should be performed again.
These are not desirable especially for large graphs.

In network epidemiology, the reduction of edges is utilized to prevent the spreading of disease by restricting connections between regions, e.g., points-of-interest and cities \cite{nowzari2016, nishi2020, chang2021a}.
However, the reduction algorithms are often ad-hoc, and their theoretical guarantee is limited.

\section{Edge Sampling Using Graph Sampling Theory}\label{sec: edge_sampling}
In this section, we describe our proposed edge sampling method.
First, the formulation and overview of our method are presented, then the details of the selection algorithm are discussed.

\begin{figure}[t]
    \centering
    \subfloat[Histogram of $(\lbrack\Wm\rbrack_{ij} - \lbrack\Wm\rbrack_{ik})^2$]{\includegraphics[width=0.69\linewidth]{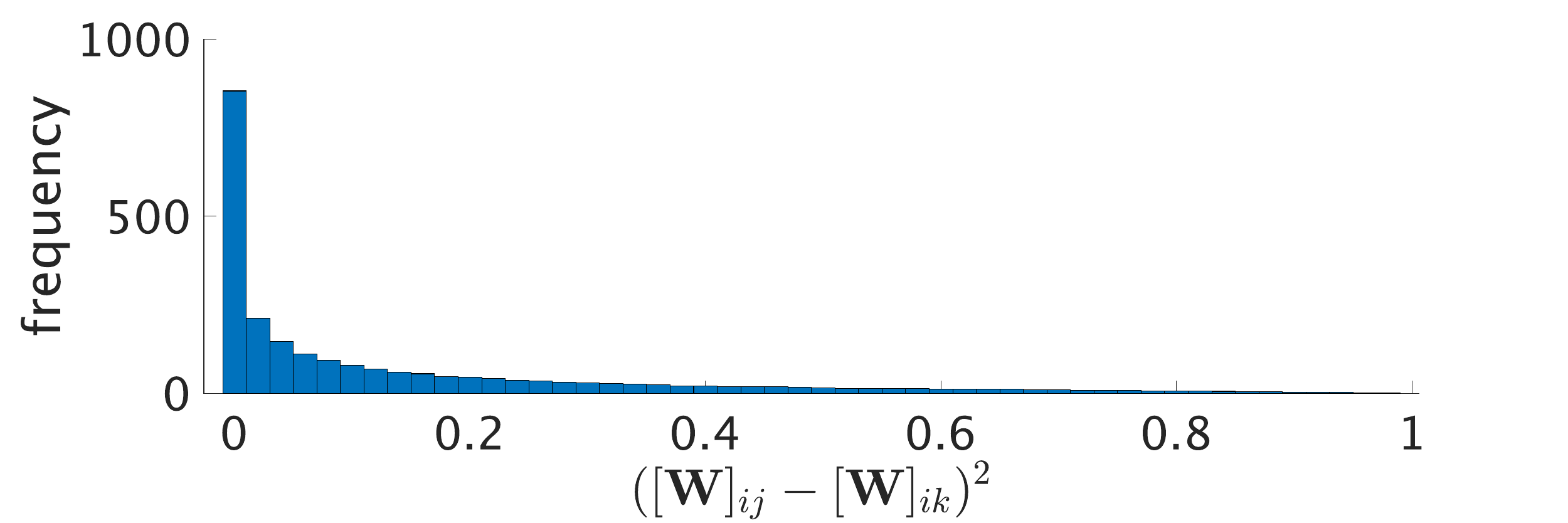}}
    \subfloat[Example of graphs]{\includegraphics[width=0.31\linewidth]{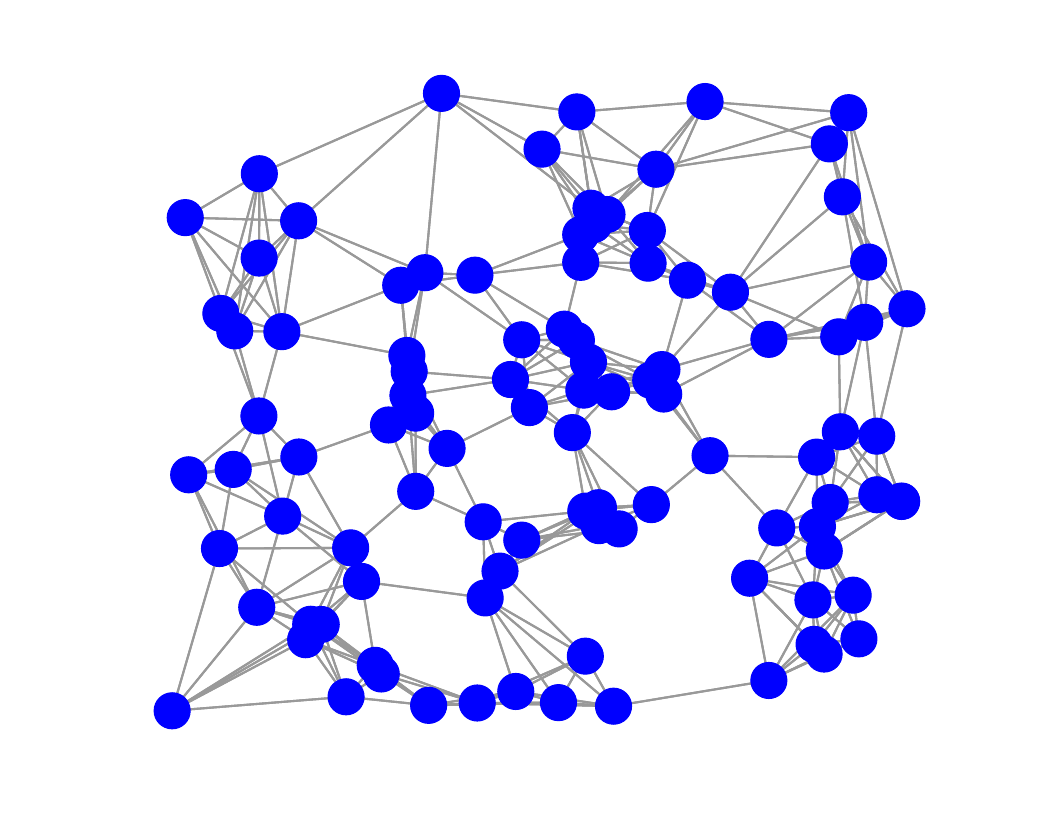}}
    \caption{(a): Histogram of the distribution of the difference between adjacent edge weights, i.e., $([\Wm]_{ij} - [\Wm]_{ik})^2$, of random sensor networks ($N = 100$). Average of 100 trials. The horizontal and vertical axes represent the difference of the edge weights and their frequency, respectively. (b): Example of a random sensor network used.}
    \label{fig:smoothness}
\end{figure}

\subsection{Formulation and Assumption}
Suppose that the original graph $\Gc_0 = (\Vc, \Ec)$ is given.
In general, edge sampling can be represented as the following objective function:
\begin{equation}
\label{eqn: prob}
    \text{find } \Fc \subset \Ec \text{ such that } \min f(\Gc_1),
\end{equation}
where $\Gc_1 := (\Vc, \Fc)$ and $f(\Gc_1)$ is some cost function.
Note that \eqref{eqn: prob} is in general NP-hard.
We thus consider solving \eqref{eqn: prob} approximately with a GSP technique.

In this paper, we consider the following cost function:
\begin{equation}
\label{eqn: costfunc}
    f(\Gc_1) := \|\wv - \text{Interp}(\wv_{\Fc} + \nv)\|_2,
\end{equation}
where $\wv \in \mathbb{R}^{|\Ec|}$ is a vector whose $[\wv]_\alpha$ is the edge weight of edge $\alpha$,
$\wv_{\Fc}$ is the subvector which has the elements of $\wv$ specified by $\Fc$, $\nv$ is i.i.d.~white Gaussian noise, and $\text{Interp}$ is a (linear) interpolation function corresponding to the chosen edge sampling method.
That is, \eqref{eqn: costfunc} leads to finding a good edge subset so that it well estimates removed edge weights under the noisy condition.
Furthermore, edge weights in a physical measurement are often perturbed or unstable during the sensing process, such as sensor networks and biomedical information processing \cite{schaub2018, barbarossa2020, Ortega2018}.
Therefore, we consider the robustness against edge weight perturbation.

In this paper, we assume the smoothness of edge weights in $\Ec$.
This can be formulated as follows:
\begin{equation}
\label{eqn: edgesmoothness}
    \sum_{i\in\Vc} 
    \sum_{j,k \in \Nc_i} \left([\Wm]_{ij} - [\Wm]_{ik}\right)^2 \le \epsilon_e,
\end{equation}
where $\Nc_i$ is the neighborhood of the node $i$ and $\epsilon_e$ is a small constant.
\eqref{eqn: edgesmoothness} means the variation of the edge weights for adjacent edges is small.
We call this \textit{edge smoothness} in this paper.

\subsubsection{Edge Smoothness for Unweighted Case}
Simply, edge smoothness can be verified through unweighted graphs.
An unweighted graph has binary edge weights $\{0, 1\}$, and therefore, $[\Wm]_{ij} - [\Wm]_{ik} = 0$ is always satisfied.

\subsubsection{Edge Smoothness for Weighted Case}
We then consider edge smoothness in weighted graphs.
For nodes distributed in space, we often estimate edge weights by (Euclidean) distances between nodes.
For example, if the edge weight is inversely proportional to the Euclidean distances,
the edge smoothness is approximately satisfied.

Fig. \ref{fig:smoothness}(a) shows the histogram of $([\Wm]_{ij} - [\Wm]_{ik})^2$ for random sensor graphs constructed by $k$NN ($k=6$) with $100$ nodes (Fig. \ref{fig:smoothness}(b)), where $[\Wm]_{ij} = \mathrm{exp}(-\frac{\|\pv_i-\pv_j\|_2}{0.3})$ is the edge weight determined by the coordinates of nodes $\pv_i \in \mathbb{R}^{2}$.
Edge smoothness refers to that $([\Wm]_{ij} - [\Wm]_{ik})^2$ is concentrated at the origin. 
As observed from the figure, the assumption of edge smoothness is reasonable even for weighted graphs whose weights are determined by distances.

\subsubsection{Edge Smoothness in Frequency Domain}
To introduce graph frequencies for the edge domain, we first define the GFT in the edge domain.
By using the right singular vector matrix $\Vm$ obtained from the singular value decomposition $\bar{\Bm} = \Um\bm{\Sigma}\Vm^\top$, the GFT of the edge domain is defined as follows:
\begin{equation}
    \hat{\wv} = \Vm^\top \wv,
\end{equation}
where $\hat{\wv}$ is the graph Fourier coefficients.
Note that, as in the definition of eigenvalue decomposition, the singular values are ordered in ascending order as $\sigma_0<\sigma_1\leq\sigma_2\cdots\leq\sigma_{N-1}=\sigma_{\mathrm{max}}$ and the corresponding singular vectors are sorted accordingly.
Similar to smoothness in the node domain, smoothness in the edge domain can be viewed as the energy of $\hat{\wv}$ being dominant at the small singular value of $\bar{\Bm}$ \cite{schaub2018, schaub2021}.
Therefore, we utilize the established definition of \textit{signal smoothness} on graphs \cite{Chen2015, Anis2016, Tanaka2020c} for \textit{edge smoothness} as follows:
\begin{equation}
\label{eqn: signalsmoothness}
    \sum_{\alpha=K}^{|\Ec|-1} \hat{w}^2_\alpha \le \epsilon_s,
\end{equation}
where $K$ is the bandwidth of $\wv$ and $\epsilon_s$ is a small constant.

Based on the above assumptions and observations, if we can \textit{flip} roles of the nodes and edges, we can use a node sampling method to select important edges.
In the following, we describe our graph conversion and edge selection method.

\begin{figure}[t]
    \centering
    \includegraphics[width = 0.8\linewidth]{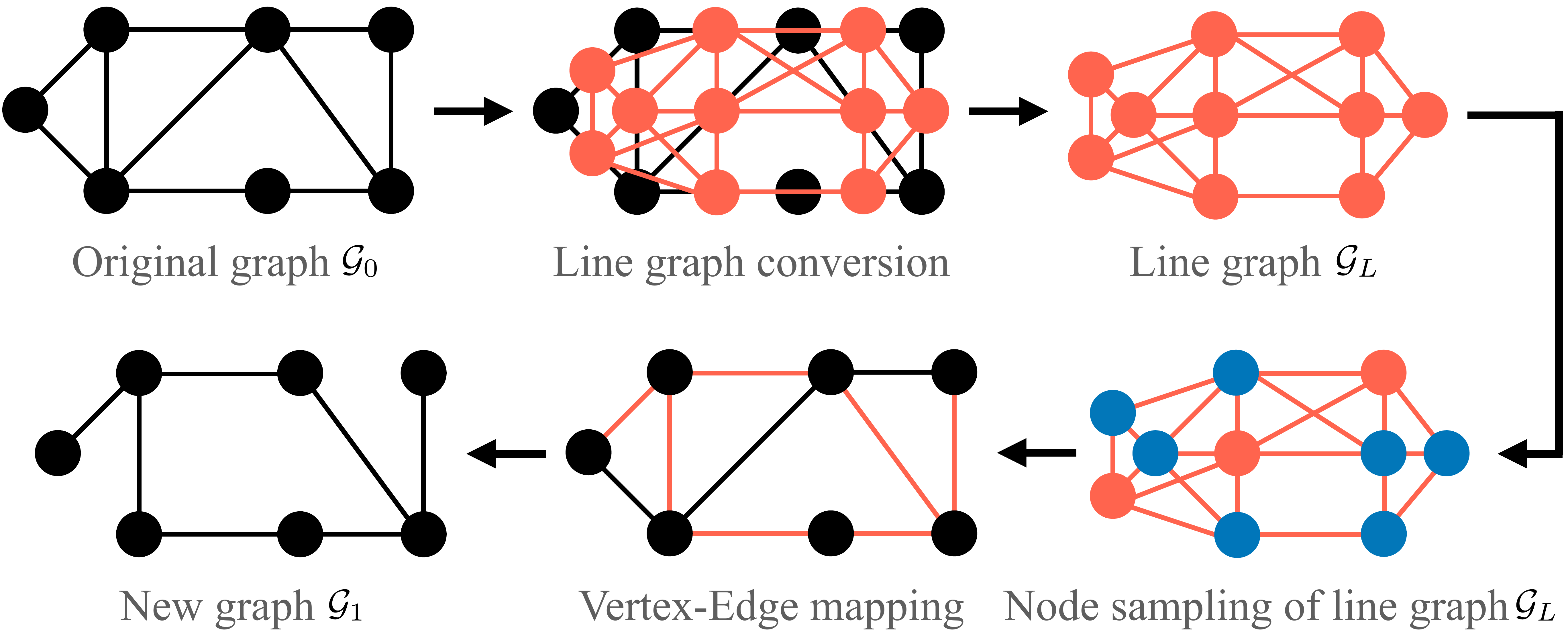}
    \caption{Overview of the proposed method. The original graph is first converted into the line graph (red nodes correspond to the original edges). The important nodes (blue) are then selected from the line graph: This extracts the most important edges in the original graph.}
    \label{fig:proposed_method}
\end{figure}

\subsection{Framework}
The overview of the proposed edge sampling method is illustrated in Fig. \ref{fig:proposed_method}.
In contrast to existing edge sparsification and reduction approaches, we first convert the original graph $\Gc_0$ into a \textit{line graph} $\Gc_L$ \cite{evans2010} and then perform sampling set selection on nodes of the line graph.
A node in the line graph represents an edge in the original graph, and the edge weight of the line graph indicates the relationship between neighboring edges in the original graph.
Therefore, \textit{the node selection of the line graph can be regarded as the edge selection of the original graph}.

Since the nodes in the line graph refer to the original edges, the edge weight of the edge $\alpha$ can be regarded as the signal value on the node $\alpha$ of $\Gc_L$.

\subsection{Graph Conversion}\label{sec: graph_conversion}
The original graph and its converted version have a concrete relationship.
First of all, we formally define the line graph.

\begin{definition}[Line graph]
Suppose that the original graph $\Gc_0 = (\Vc, \Ec)$ and its incidence matrix $\tilde{\Bm}$ are given.
The adjacency matrix of the line graph $\Wm_L \in \mathbb{R}^{|\Ec| \times |\Ec|}$ is defined as follows \cite{evans2010}:
\begin{equation}
\label{eqn: adj_line}
\begin{split}
[\Wm_{L}]_{\alpha\beta} &= \sum_i [\tilde{\Bm}^\top]_{\alpha i}[\tilde{\Bm}]_{i\beta}(1-\delta_{\alpha\beta}) \\&= \sum_i [\tilde{\Bm}^{\top}]_{\alpha i}[\tilde{\Bm}]_{i\beta} - 2[\Cm]_{\alpha\beta},
\end{split}
\end{equation}
where $\alpha$ and $\beta$ are edge indices of the original graph (and therefore, they are node indices in $\Gc_{L}$), $\delta_{\alpha\beta}$ is Kronecker delta, and $\Cm:= \emph{diag}(\wv)$ in which $\wv \in \mathbb{R}^{|\Ec|}$ is the edge weight vector sorted by the edge indices.
For unweighted graphs, the line graph is also obtained using \eqref{eqn: adj_line} by replacing $\tilde{\Bm}$ and $\Cm$ with $\Bm$ and the identity matrix $\Id_{|\Ec|}$, respectively.
\end{definition}
\noindent

The graph Laplacian $\Lm_L$ of the line graph can be introduced as $\Lm_L = \Dm_L - \Wm_L$ where $\Dm_L$ is the degree matrix of $\Gc_L$.
In addition to the graph Laplacian $\Lm_L$
, we also use \textit{edge Laplacian} as a possible operator for signed line graphs, where directed edge weights can also be considered.
\begin{definition}[Edge Laplacian]
\label{def: Edge laplacian}
Suppose a directed incidence matrix $\bar{\Bm}$ in \eqref{eqn: incidence_directed} is given.
The edge Laplacian is defined as follows \cite{roddenberry2022}:
\begin{equation}
\label{eqn: edge_Laplacian}
    \Lm_e = \bar{\Bm}^\top\bar{\Bm}.
\end{equation}
Note that the graph Laplacian $\Lm_L$ and edge Laplacian $\Lm_e$ are different graph operators but have the same network structure.
\end{definition}

Here, we present the relationship between the degrees of the line graph and the original edge weights.
\begin{proposition}
\label{prop}
Suppose that the incidence matrix $\Bm$, $\tilde{\Bm}$, or $\bar{\Bm}$ of the original graph $\Gc_0 = (\Vc, \Ec)$ is given, and the adjacency matrix or the edge Laplacian of the line graph is given by \eqref{eqn: adj_line} or \eqref{eqn: edge_Laplacian}.
Then, the degree of the node $\alpha$ in the line graph, $d_{\alpha}$, that corresponds to the edge $\alpha$ connecting the nodes $m$ and $n$ in the original graph is obtained as follows.
\begin{equation}
\label{eqn: weight_line_graph}
    d_{\alpha} = 
    \begin{cases}
    k_m + k_n - 2 & \text{for unweighted graphs},\\
    \sqrt{w_\alpha} (\tilde{d}_m + \tilde{d}_n) - 2w_{\alpha} & \text{for weighted graphs},\\
    \sqrt{w_\alpha} (\bar{d}_m - \bar{d}_n) & \text{for directed graphs},
    \end{cases}
\end{equation}
where $w_{\alpha}$ is the weight of the edge $\alpha=(m,n)$, $k_m$ is the number of edges connecting to the node $m$, and $\tilde{d}_m$ and $\bar{d}_m$ are the weighted degree of the node $m$ calculated by $\sum_n \sqrt{w_{mn}}$.

\end{proposition}
\noindent

\begin{proof}
The degree of the node $i$ of the original graph $\tilde{d}_i$ can be expressed using the incidence matrix $\tilde{\Bm}$ as follows.
\begin{equation}
    \tilde{d}_i = \sum_{\alpha}[\tilde{\Bm}]_{i\alpha},
\end{equation}
where $\alpha$ is the edge index.
Furthermore, the degree $d_{\alpha}$ of the node $\alpha$ on the line graph is given by
\begin{equation}
\label{eqn: degree}
    d_{\alpha} = \sum_{\beta}[\Wm_{L}]_{\alpha\beta}.
\end{equation}
\eqref{eqn: degree} can be further rewritten as follows:
\begin{equation}
\label{eqn: kalpha}
    \begin{split}
        d_{\alpha} = & \sum_{\beta}[\Wm_{L}]_{\alpha\beta} \\
         = & \sum_{\beta}\left(\sum_i [\tilde{\Bm}^\top]_{\alpha i}[\tilde{\Bm}]_{i\beta} - 2[\Cm]_{\alpha\beta}\right)\\
         = & \sum_{\beta}\sum_i [\tilde{\Bm}^\top]_{\alpha i}[\tilde{\Bm}]_{i\beta} - 2\sum_{\beta}[\Cm]_{\alpha\beta}\\
         = & \sum_{i}[\tilde{\Bm}^\top]_{\alpha i}\sum_{\beta}[\tilde{\Bm}]_{i\beta} - 2w_{\alpha}\\
         = & \sum_{i}\tilde{d}_{i}[\tilde{\Bm}^\top]_{\alpha i} - 2w_{\alpha}.
    \end{split}
\end{equation}

Since $m$ and $n$ are the nodes connected to the edge $\alpha$, \eqref{eqn: kalpha} is rewritten with $w_{\alpha}$ as follows:
\begin{equation}
    \begin{split}
        d_{\alpha} & = \sqrt{w_\alpha} \tilde{d}_m + \sqrt{w_\alpha} \tilde{d}_n - 2w_{\alpha}\\
        & = \sqrt{w_{\alpha}}(\tilde{d}_m + \tilde{d}_n) - 2w_\alpha.
    \end{split}
\end{equation}
This is identical with \eqref{eqn: weight_line_graph} for weighted graphs.

If the original graph $\Gc_0$ is an unweighted graph without a pseudo-orientation, then $w_{\alpha} = 1$ and $\tilde{d}_m = k_m$.
Hence, the degree of the edge $\alpha$ is $d_{\alpha} = k_m  + k_n - 2$.

When $\Gc_0$ is a directed graph (or introduces a pseudo-orientation), the line graph is represented by the edge Laplacian. 
In this case, the line graph is derived from $\bar{\Bm}^\top\bar{\Bm}$ instead of $\tilde{\Bm}^\top\tilde{\Bm}-2\Cm$: The second term in \eqref{eqn: kalpha} becomes 0.

In addition, as mentioned in the Notation, $\bar{\Bm}$ has positive and negative elements in each column; thus, the degree of the edge $\alpha$ is $d_\alpha = \sqrt{w_\alpha} (\bar{d}_m - \bar{d}_n)$.
This completes the proof.
\end{proof}

Proposition \ref{prop} indicates that the high-degree nodes in the line graph correspond to the original edges connecting high-degree nodes.
In sampling set selection based on signal smoothness, selected nodes often have high degrees.
Hence, selecting important nodes in the line graph can be regarded as selecting important edges in the original graph.

\begin{table*}[t]
    \caption{Computational Complexities of edge sampling methods.
    Parameters: $|\Fc|$: Number of sampling edges; $P$: Approximation order of the Chebyshev polynomial approximation; $J$: Number of non-zero elements of the localization operator\cite{perraudin2018}.}
    \centering
    \begin{tabular}{c|c|c}\hline
        & Preparation & Selection \\ \hline\hline
      Proposed & $\Oc (|\Ec|^2 + p|\Ec||\Ec_L| + J)$ & $\Oc(J|\Fc|)$ \\ \hline
      Proposed-Faster & $\Oc (|\Ec|^2 + pN|\Ec| + J)$ & $\Oc(J|\Fc|)$ \\ \hline
      NetMelt\cite{chen2016} & $\Oc(N + |\Ec|)$ & $\Oc (|\Ec||\Fc|)$ \\ \hline
      MaxDegree & $\Oc(N|\Ec|)$ & $\Oc(|\Ec|\log |\Ec|)$ \\\hline
      GSparse \cite{spielman2011} & $\Oc (\log N)$ & $\Oc(|\Ec|)$ \\ \hline
    \end{tabular}
    \label{tbl:computational_complexity}
\end{table*}

\subsection{Node Sampling of Line Graph}
As previously mentioned, we can use an arbitrary sampling set selection for the line graph.
The important property of GSP-based sampling set selection methods is that many of them are designed to be robust to noise \cite{Tanaka2020c}.
This satisfies the requirement of \eqref{eqn: costfunc}.

While any sampling set selection algorithm can be utilized, two issues should be considered according to applications.
The first property is the distribution of the sampled edges.
Especially for sensor networks, selected edges are desired to be distributed uniformly in space.
That is, distributed selection algorithms are beneficial as edge sparsification.
In contrast, concentrated selection algorithms do not have such a restriction on the node distribution.
This could be utilized to prevent the spread of infections in the context of network epidemics.

The second property is computation complexity.
The number of nodes in the line graph is $|\Ec|$, which is often greater than that of the original graph $N$.
Hence, fast selection methods are preferred, especially for edge sampling.

\section{Accelerated Node Sampling of Line Graph}\label{sec: acceleration}
In this section, we consider graph sparsification as an application of edge sampling.
Here, we consider the FastGSSS\cite{sakiyama2019} as a fast node sampling method for the proposed method since the sampling nodes are spatially uniform and appropriate for graph sparsification.

FastGSSS does not require the eigendecomposition of graph operators by using a Chebyshev polynomial approximation (CPA) of the filter kernel.
However, we still need a high computation cost for edge sampling even if we use CPA because it requires the matrix multiplications of size $|\Ec|\times|\Ec|$.
In the following, we describe the further acceleration method of FastGSSS for edge sampling.

\subsection{Acceleration by Filtering on Original Graph}\label{sec: acc_filter}
FastGSSS selects a node $\alpha$ of $\Gc_L$ by repeating $|\Fc|$ iterations of the following equation\cite{sakiyama2019}:
\begin{equation}
    \alpha^* = \underset{\alpha \in \Sc_m^c}{\arg \max } \left\langle R\left(\eta \mathbf{1}_{|\Ec|} - \sum_{\beta \in \Sc_m}\left|\boldsymbol{T}_{g, \beta}\right|\right), \left|\boldsymbol{T}_{g, \alpha}\right| \right\rangle,
\end{equation}
where $\boldsymbol{T}_{g, \alpha}$ is the $\alpha$th column of the localization operator $\Tm$ with the spectral kernel $g(\cdot)$, $\Sc_m$ and $\Sc_m^c$ are the selected nodes in the $m$th iteration and its rest of nodes $\Vc\backslash\Sc_m$, $\eta$ is an aribtrary positive value, and $R(\cdot)$ is the ramp function satisfied with $[R(\xv)]_\alpha = [\xv]_\alpha$ if $[\xv]_\alpha \geq 0$ and $0$ otherwise.

Let $g(x)$ be the spectral kernel; then, the localization operator $\Tm$ of the line graph is defined as follows\cite{perraudin2018a}:
\begin{equation}
    \Tm = \sqrt{|\Ec|}g(\Lm_e) = \sqrt{|\Ec|}g(\bar{\Bm}^\top\bar{\Bm}) = \sqrt{|\Ec|} \Vm g(\bm{\Lambda}_e) \Vm^\top,
    \label{eqn: Localization_operator}
\end{equation}
where $\bm{\Lambda}_e$ and $\Vm$ are the eigenvalue matrix of $\Lm_e$ and its corresponding eigenvector matrix (also singular vector matrix of $\bar{\Bm}$), respectively.

Even when simply using the graph Laplacian $\Lm_L$ of $\Gc_L$, fast edge sampling can be achieved by using fast node sampling methods such as FastGSSS that do not require eigenvalue decomposition.
Further acceleration of edge sampling is also possible when the edge Laplacian is used in the proposed framework.

The graph Laplacian of the original graph and the edge Laplacian of the line graph have the same nonzero eigenvalues\cite{barbarossa2020}.
In other words, the filter response of the $\Lm$ and the $\Lm_e$ are identical, and we assume that filtering in the edge domain can be approximated by filtering in the node domain of the original graph.
The following method focuses on this relationship and designs $g'(\cdot)$ such that $\bar{\Bm}^\top g'(\Lm)\bar{\Bm}$ approximates $g(\Lm_e)$.

Since the singular value decomposition of $\bar{\Bm}$ is $\bar{\Bm} = \Um\bm{\Sigma}\Vm^\top$, \eqref{eqn: Localization_operator} in the graph frequency domain can also be represented as follows:
\begin{equation}
\begin{split}
\Tm &= \sqrt{|\Ec|}\Vm g(\bm{\Sigma}^\top\bm{\Sigma})\Vm^\top\\
    &= \sqrt{|\Ec|}\Vm \mathrm{diag}(g(\bm{\Sigma}_N^\top\bm{\Sigma}_N), 0)\Vm^\top\\
    &= \sqrt{|\Ec|}\Vm_N g(\bm{\Lambda})\Vm_N^\top,
\end{split}
\label{eqn: Localization_operator_sigma}
\end{equation}
where $\bm{\Sigma}_N$ and $\Vm_N$ are the singular value matrix with nonzero singular values of $\bar{\Bm}$ as diagonal elements and the singular vector matrix corresponding to them.

By using the spectral kernel\footnote{$\epsilon\Id_N$ is used to ensure invertibility.} $g'(\bm{\Lambda}) = (\epsilon\Id_N+\bm{\Lambda})^{-1}g(\bm{\Lambda})$ with a small constant $\epsilon$ and the identity matrix $\Id_N$, \eqref{eqn: Localization_operator_sigma} can be approximated as follows:
\begin{equation}
\begin{split}
\Tm &= \sqrt{|\Ec|}\Vm \bm{\Sigma}^\top \bm{\Sigma}_N^{-1}g(\bm{\Lambda})\bm{\Sigma}_N^{-1}\bm{\Sigma} \Vm^\top\\
    &= \sqrt{|\Ec|}\Vm \bm{\Sigma}^\top \Um^\top \Um \bm{\Lambda}^{-1}g(\bm{\Lambda})\Um^\top\Um\bm{\Sigma} \Vm^\top\\
    &\approx \sqrt{|\Ec|}\Vm \bm{\Sigma}^\top \Um^\top \Um g'(\bm{\Lambda}) \Um^\top\Um\bm{\Sigma} \Vm^\top\\
    &= \sqrt{|\Ec|}\bar{\Bm}^\top g'(\Lm) \bar{\Bm}.
\end{split}
\label{eqn: Localization_operator_gdash}
\end{equation}
\eqref{eqn: Localization_operator_gdash} reduces computational complexity by replacing the filtering on the line graph with the filtering on the original graph.

\subsection{Computational Complexity}
In this subsection, we discuss the computational complexity of the proposed method.
Table \ref{tbl:computational_complexity} shows the computational complexity required for edge sampling when using the CPA, where $p$ is the approximate degree of the CPA, $J$ is the number of nonzero elements of the graph localization operator, $|\Fc|$ is the number of edges to sample, and $|\Ec_L|$ is the number of edges in the line graph.
Since all the methods require the preparation and selection phases, we separately compared them.

In the framework of the proposed method, the incidence matrix has only two elements in each column, thus the computation of the graph Laplacian of $\Gc_L$ requires $\Oc(|\Ec|^2)$.
Then it takes $\Oc(p|\Ec||\Ec_L|)$ to compute the localization operator.
In the case of the faster method in Section \ref{sec: acc_filter}, the computational complexity is $\Oc(|\Ec|^2+pN|\Ec|)$ because the graph Laplacian of $\Gc_0$ is used to compute the localization operator of $\Gc_L$.
In general, since $N\ll|\Ec_L|$, the computational complexity can be reduced.

\begin{figure}[t]
    \centering
    \subfloat[Sensor network]{\includegraphics[width=0.44\linewidth]{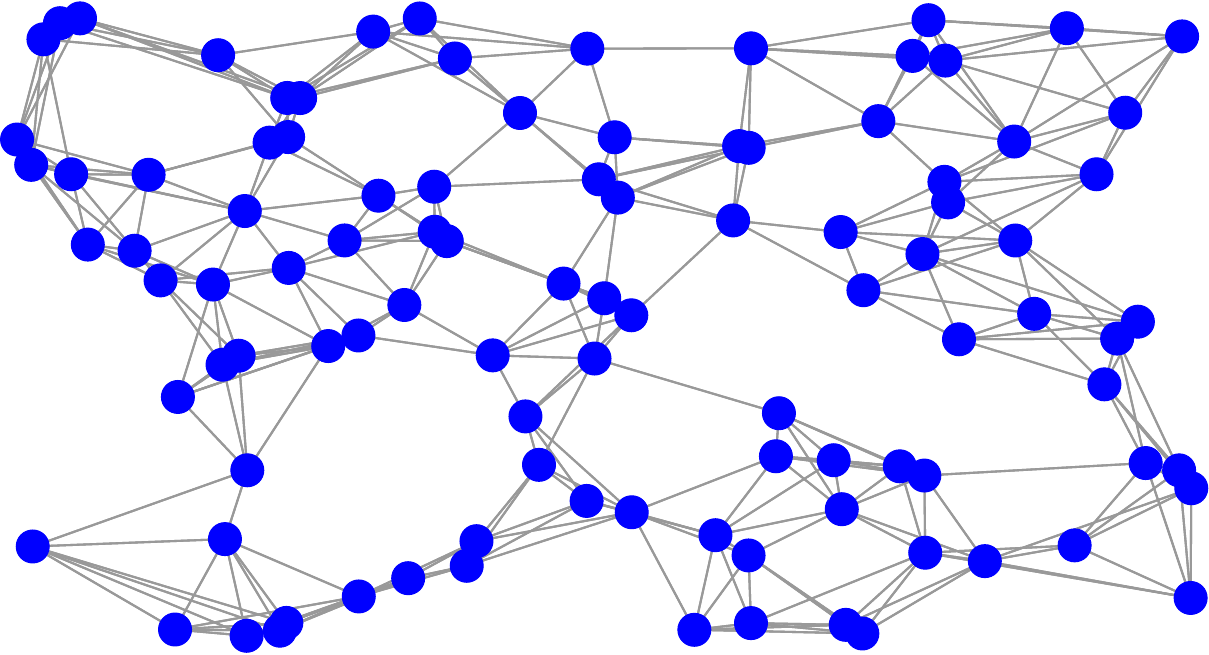}}
    \subfloat[Erd\H{o}s--R\'{e}nyi model]{\includegraphics[width=0.44\linewidth]{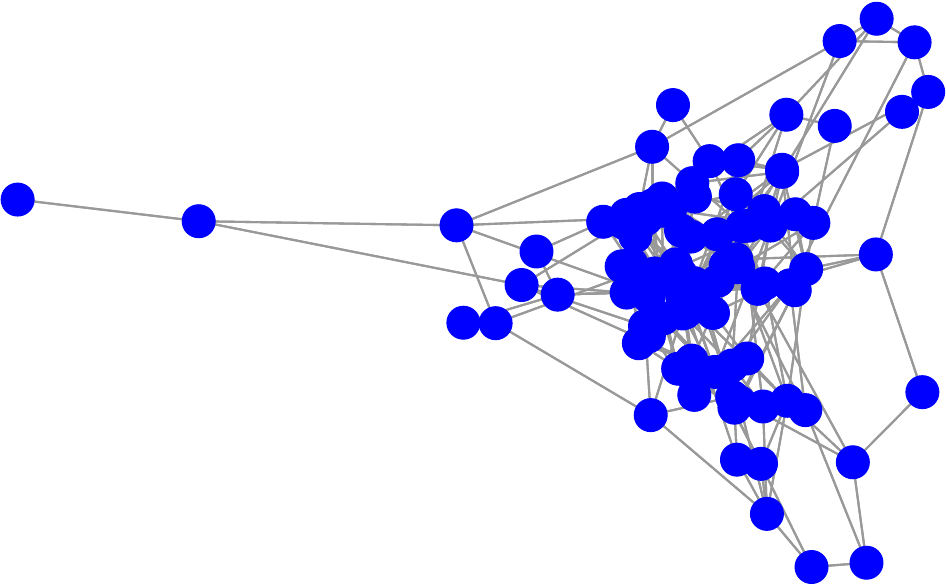}}\\
    \subfloat[Community graph]{\includegraphics[width=0.44\linewidth]{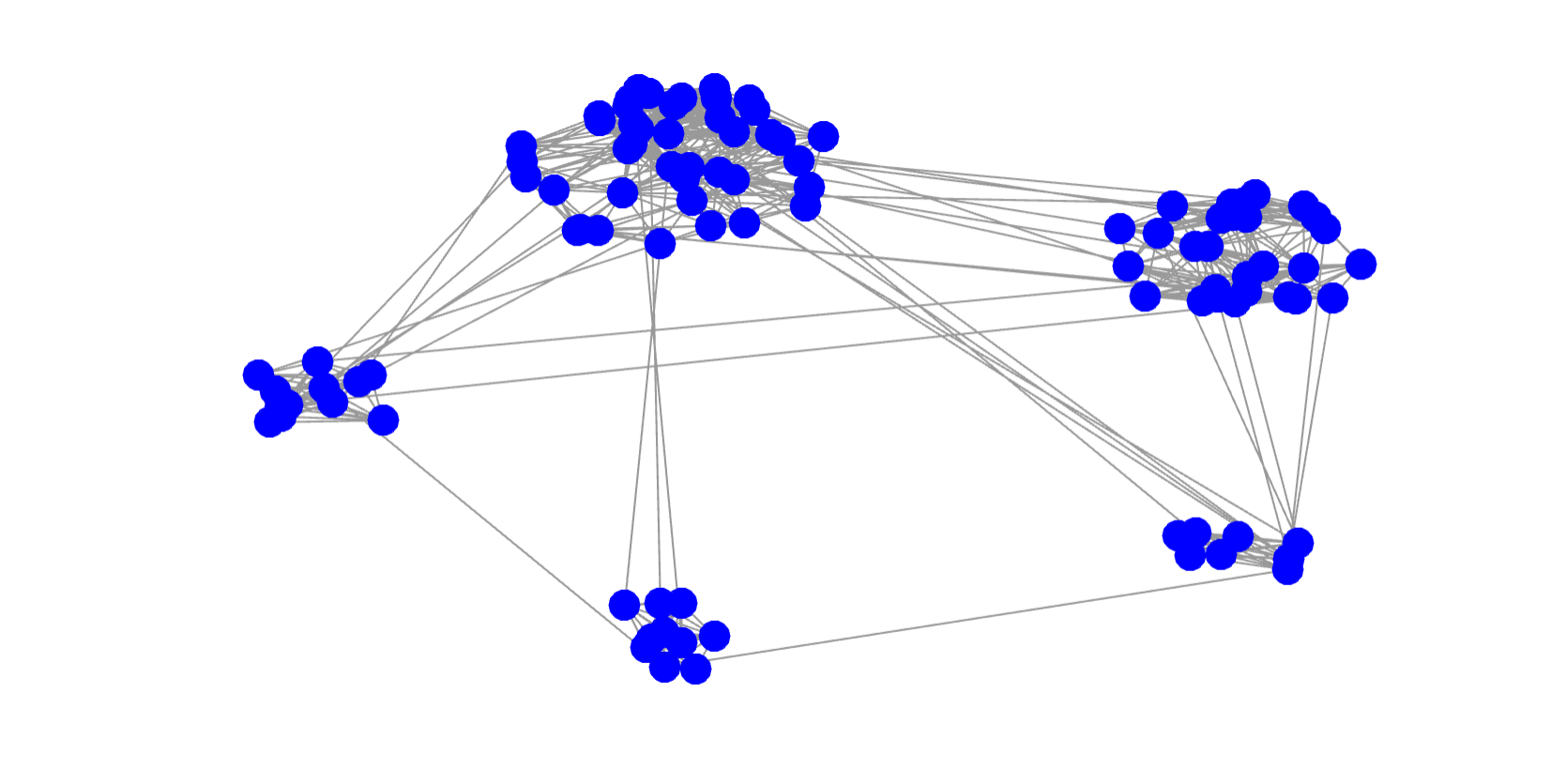}}
    \subfloat[$k$NN graph]{\includegraphics[width=0.44\linewidth]{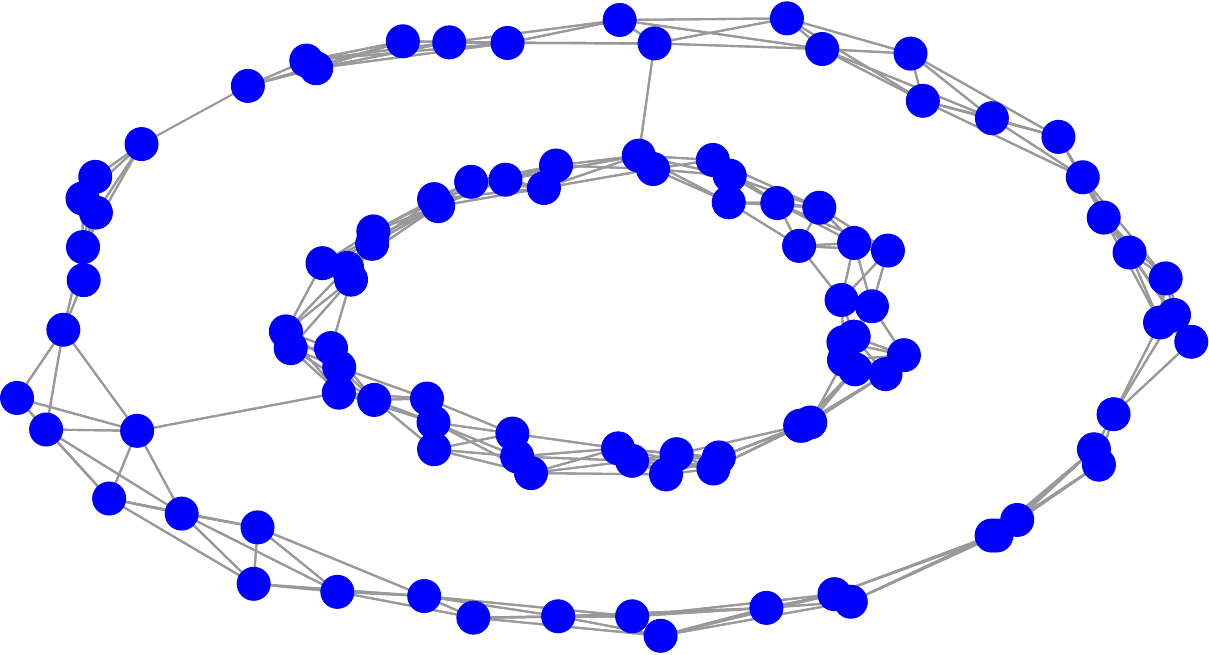}}
    \caption{Examples of graph with $N=100$.}
    \label{fig:ex-graph}
\end{figure}

\begin{figure*}[t]
\centering
\subfloat[Original graph-$\xv$]{\includegraphics[width=0.24\linewidth]{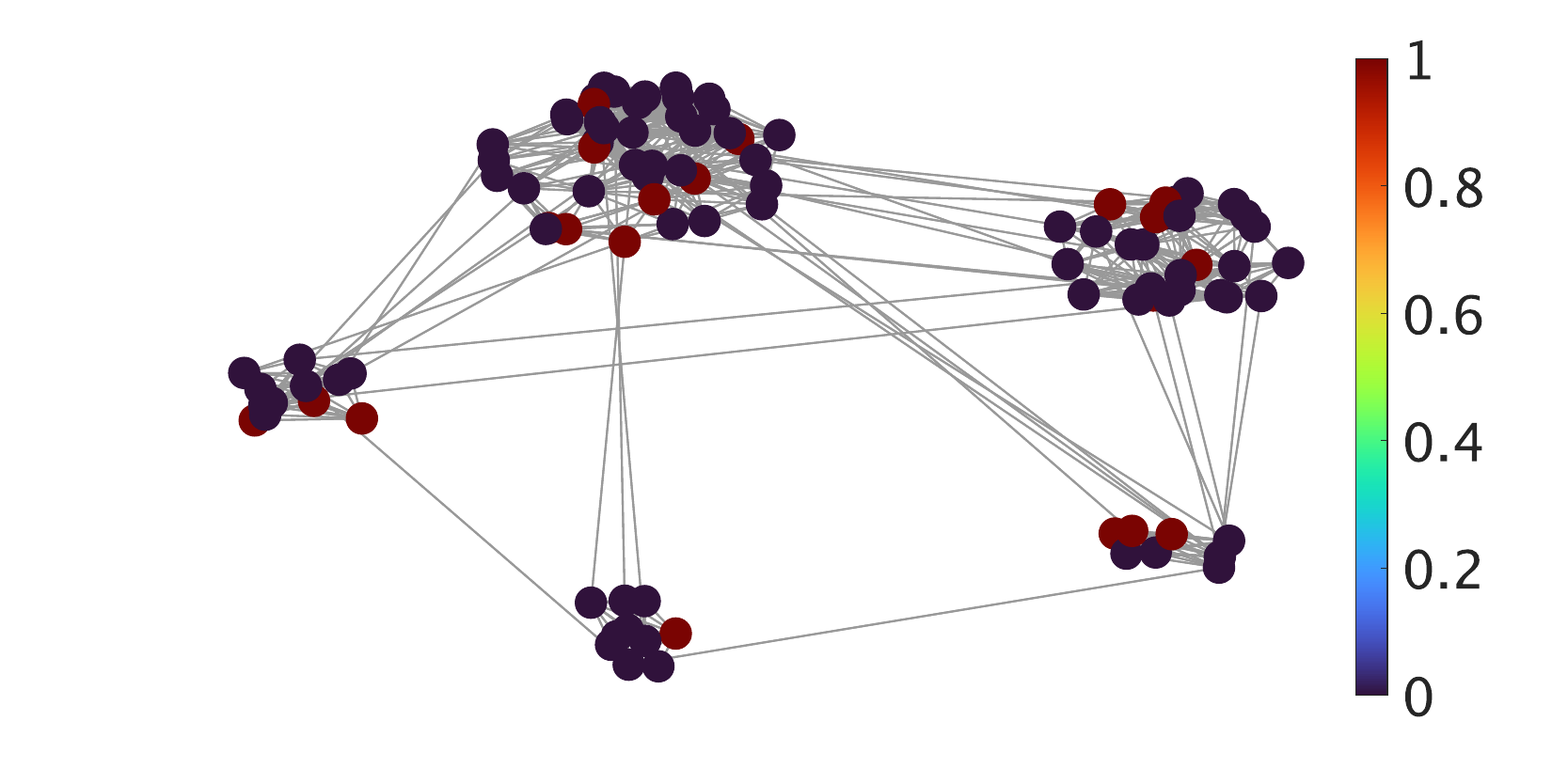}}
\subfloat[Original graph-$\yv_0$]{\includegraphics[width=0.24\linewidth]{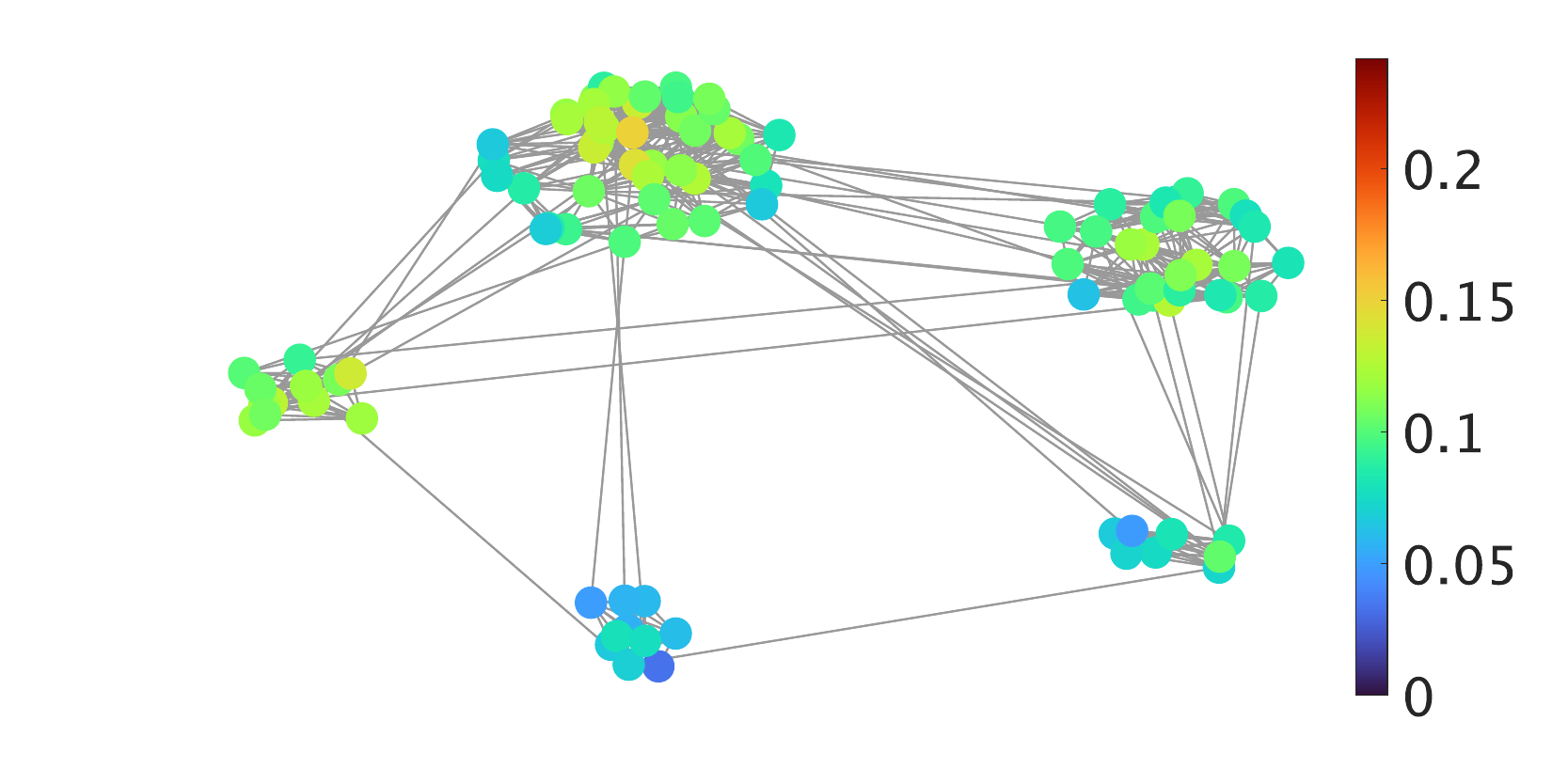}}
\subfloat[Proposed-$\yv_1$]{\includegraphics[width=0.24\linewidth]{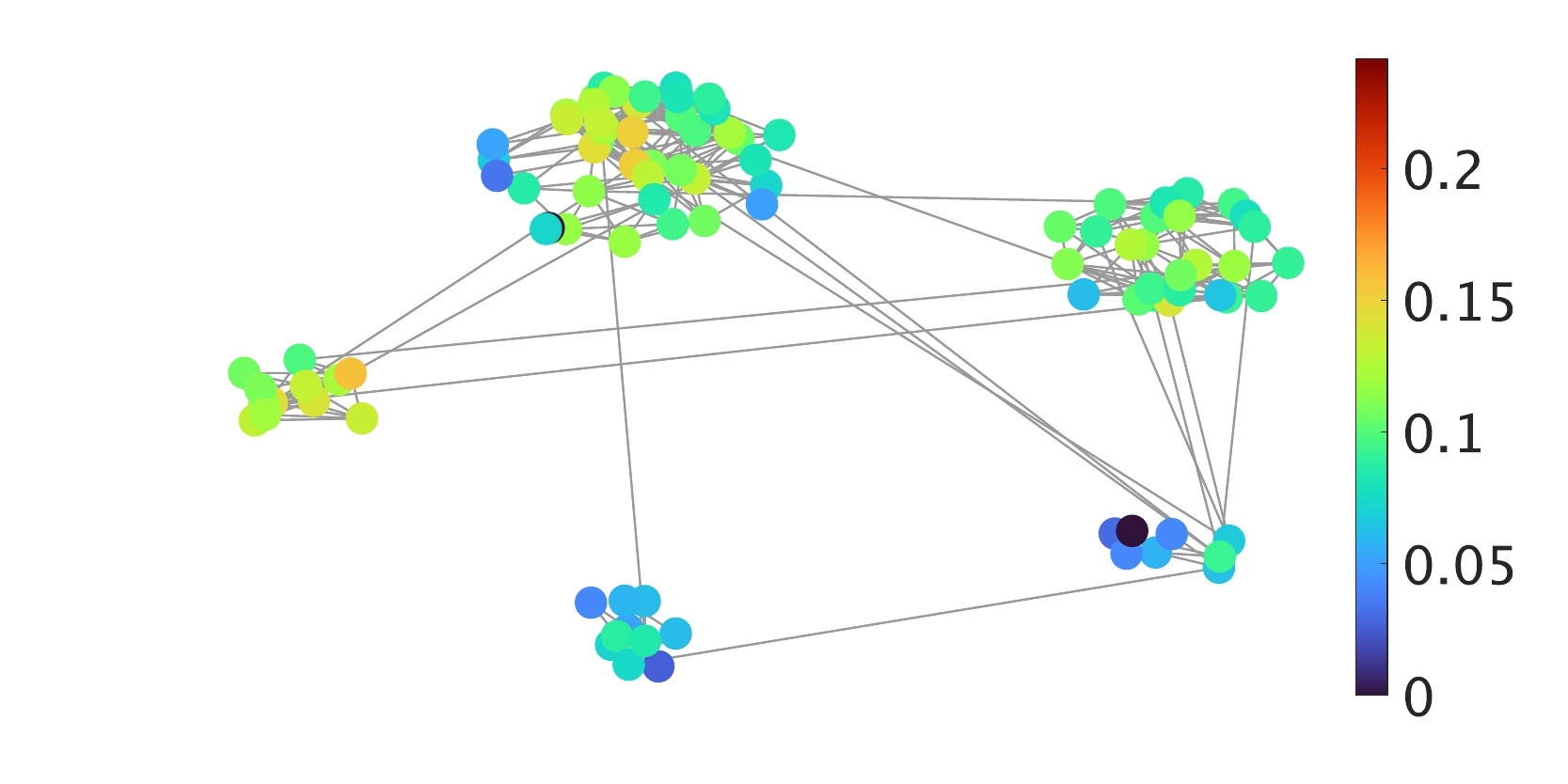}}
\subfloat[Proposed-Faster-$\yv_1$]{\includegraphics[width=0.24\linewidth]{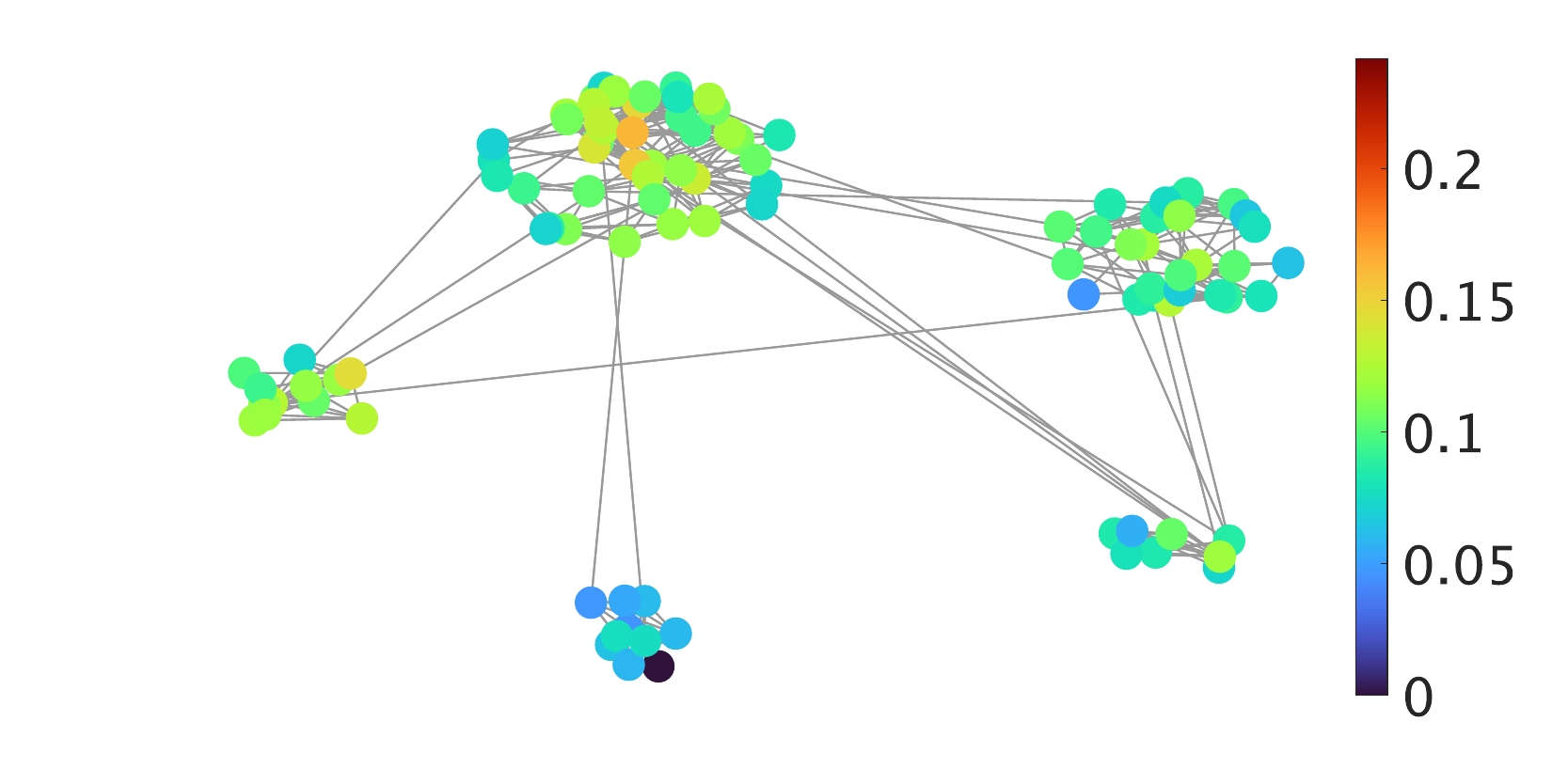}}\\
\subfloat[MaxDegree-$\yv_1$]{\includegraphics[width=0.24\linewidth]{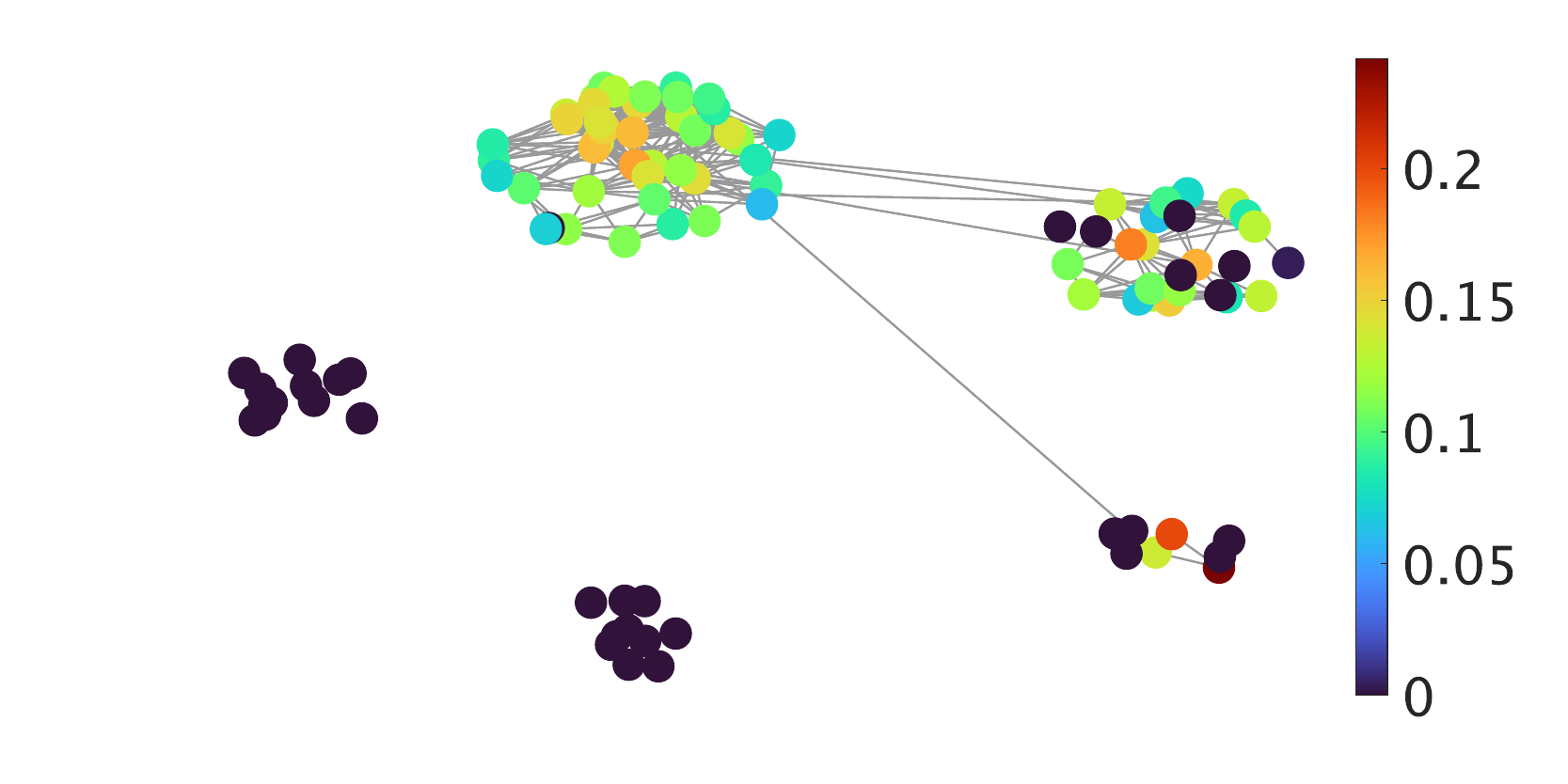}}
\subfloat[NetMelt-$\yv_1$]{\includegraphics[width=0.24\linewidth]{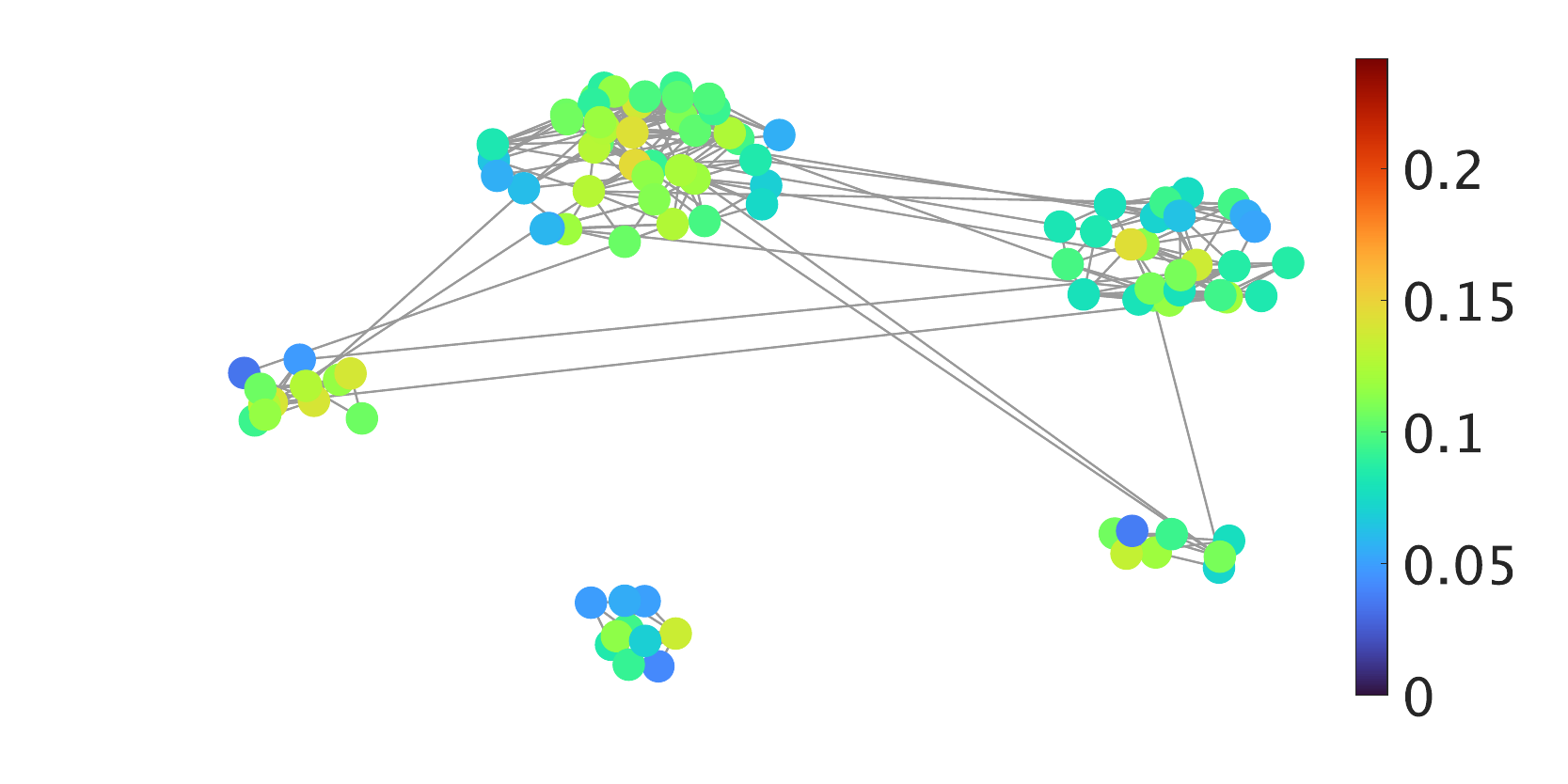}}
\subfloat[GSparse-$\yv_1$]{\includegraphics[width=0.24\linewidth]{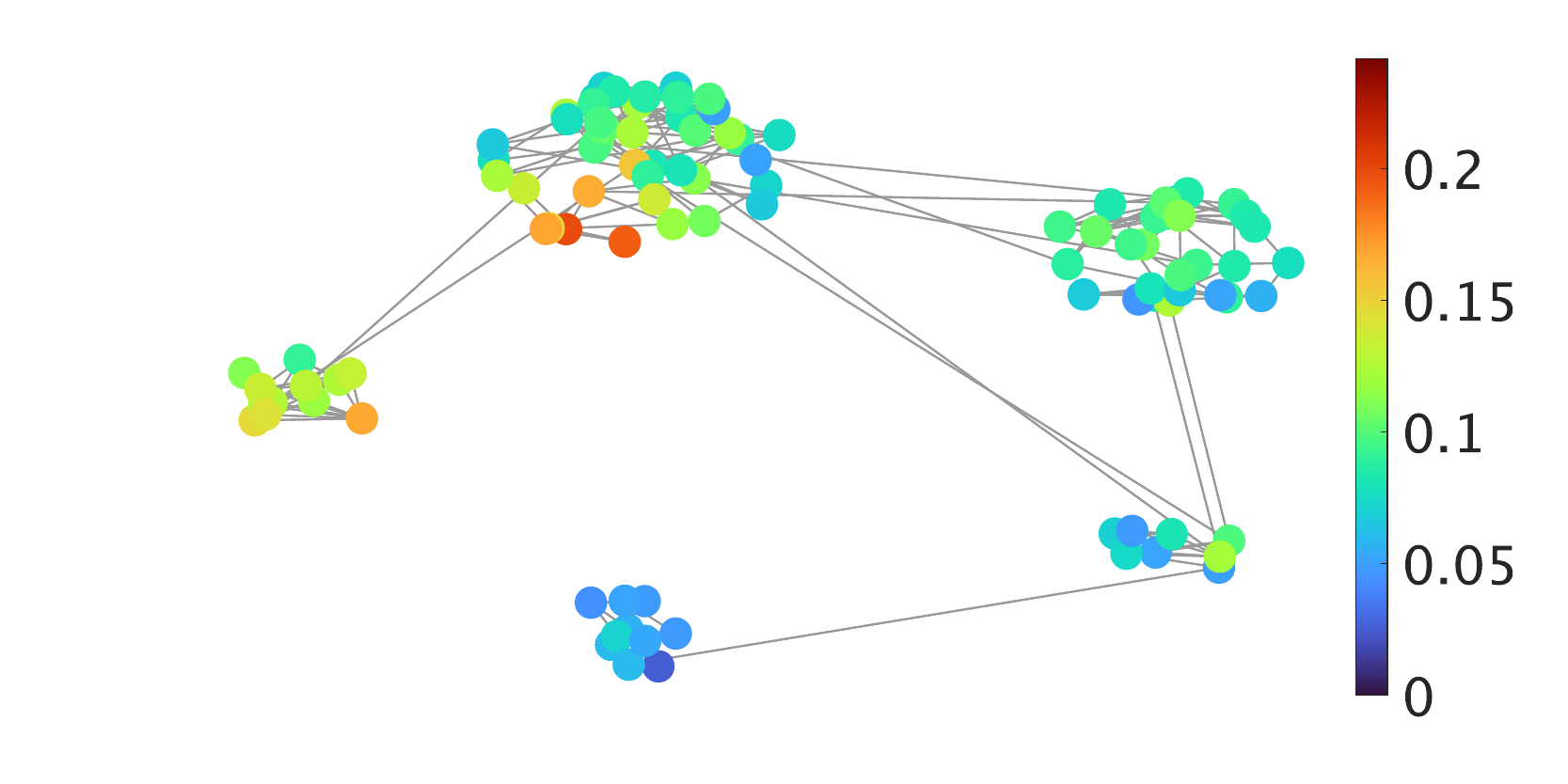}}
\caption{Graph sparsification and diffusion example: Community graph. Diffused signals are also shown.}
\label{fig:exp1-community}
\end{figure*}

\begin{figure*}[t]
    \centering
    \subfloat[Sensor network]{\includegraphics[width=0.22\linewidth]{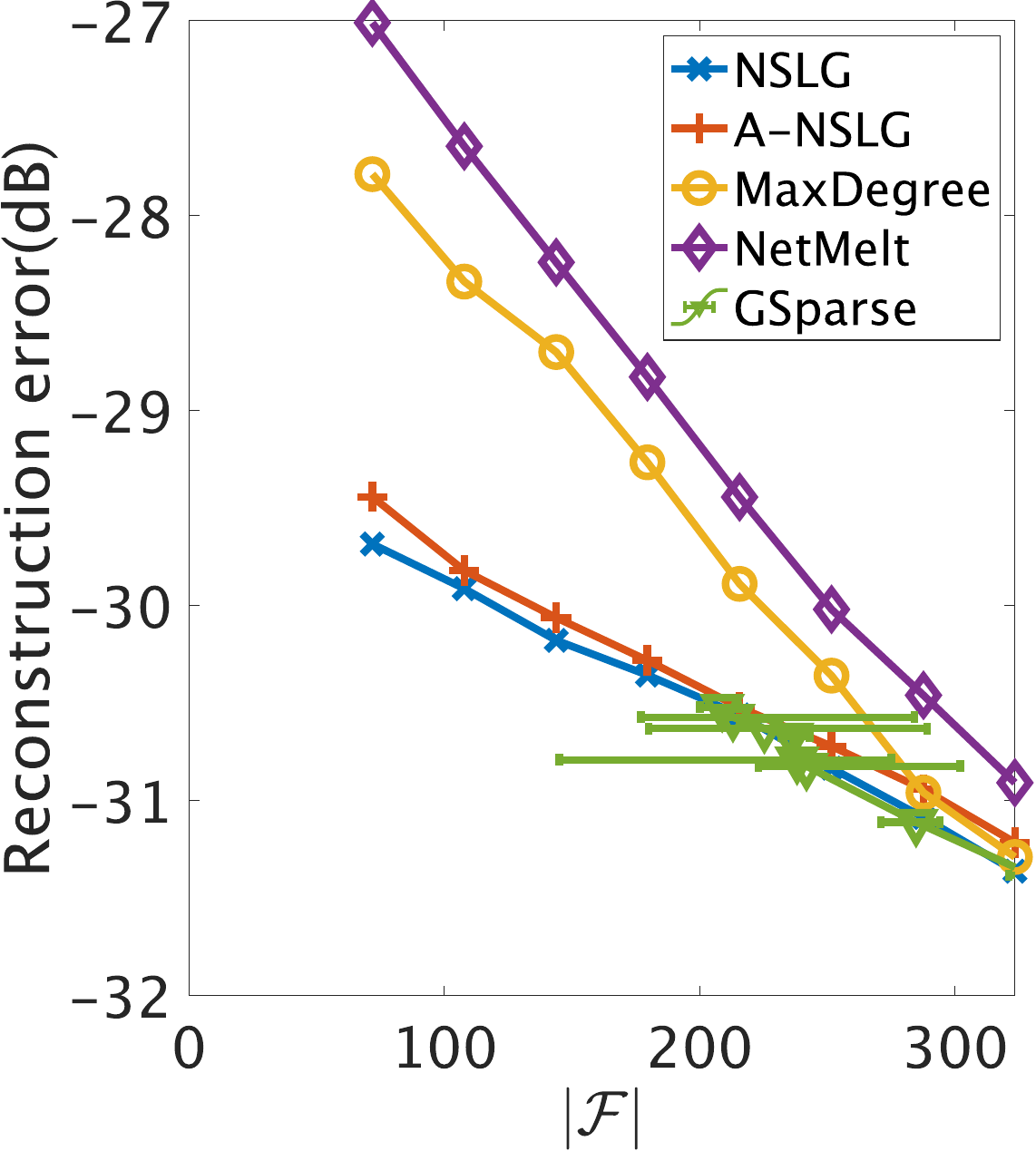}}
    \subfloat[Erd\H{o}s--R\'{e}nyi model]{\includegraphics[width=0.22\linewidth]{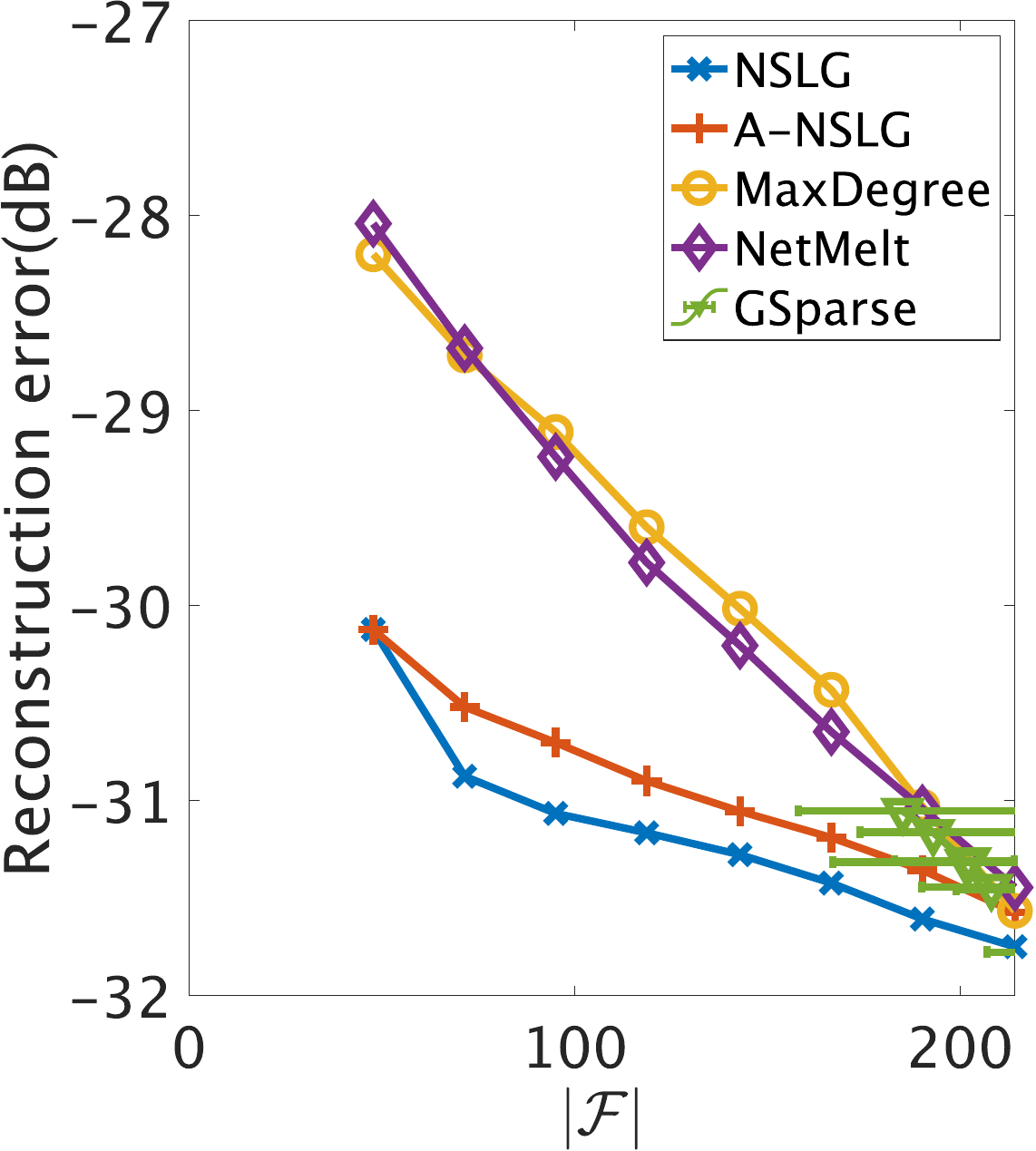}}
    \subfloat[Community graph]{\includegraphics[width=0.22\linewidth]{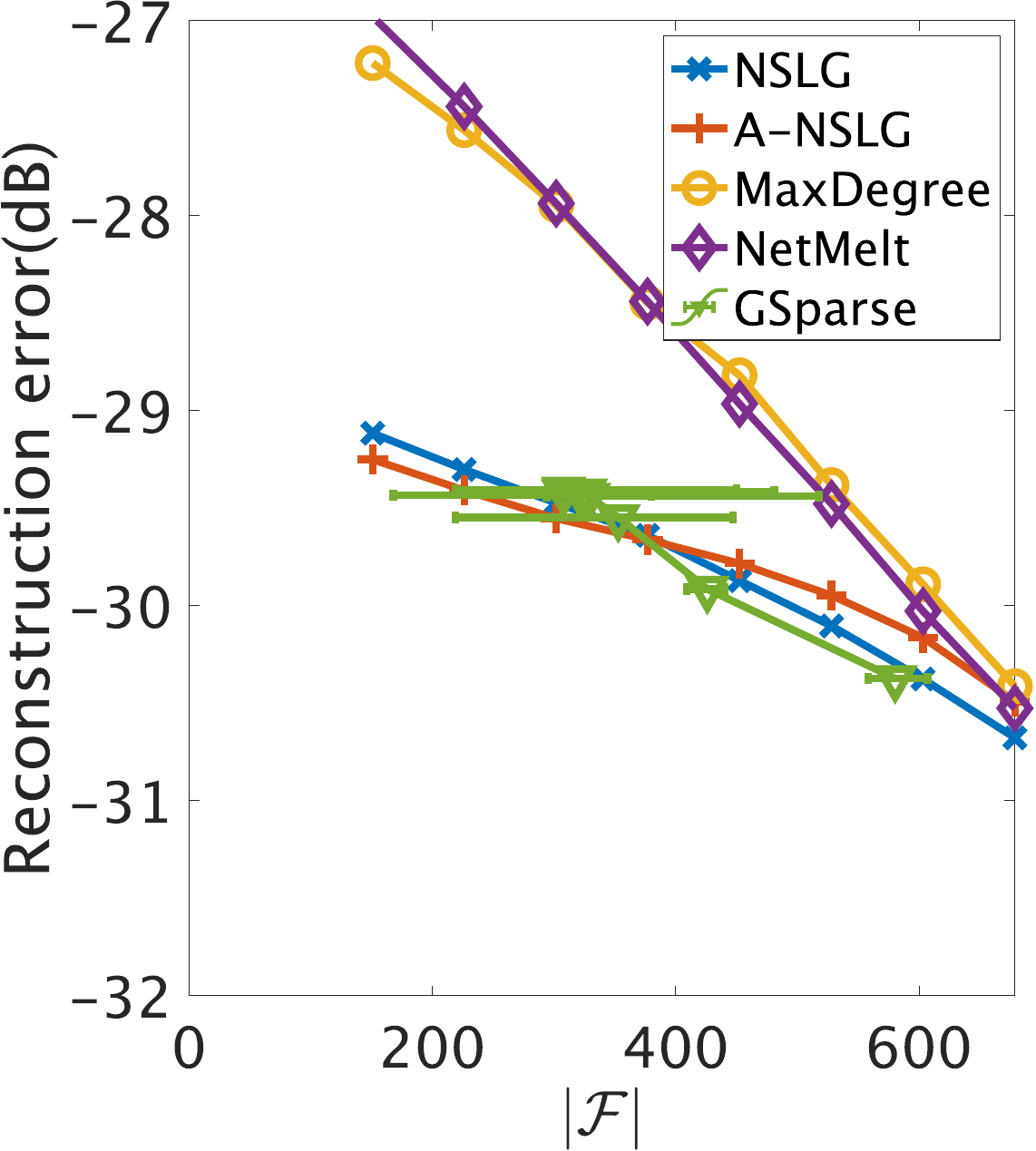}}
    \subfloat[$k$NN graph]{\includegraphics[width=0.22\linewidth]{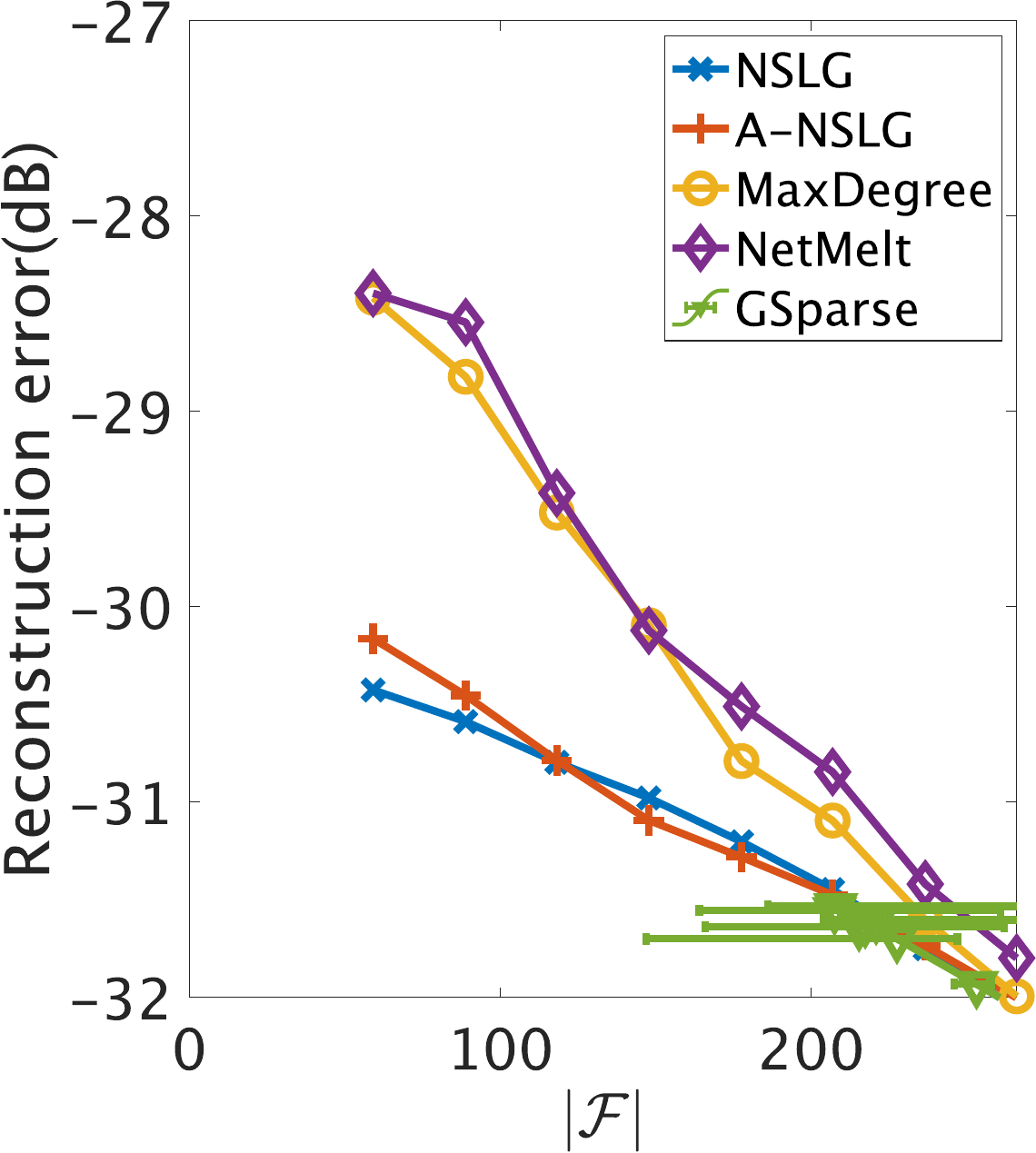}}\\
    \subfloat[Sensor network]{\includegraphics[width=0.22\linewidth]{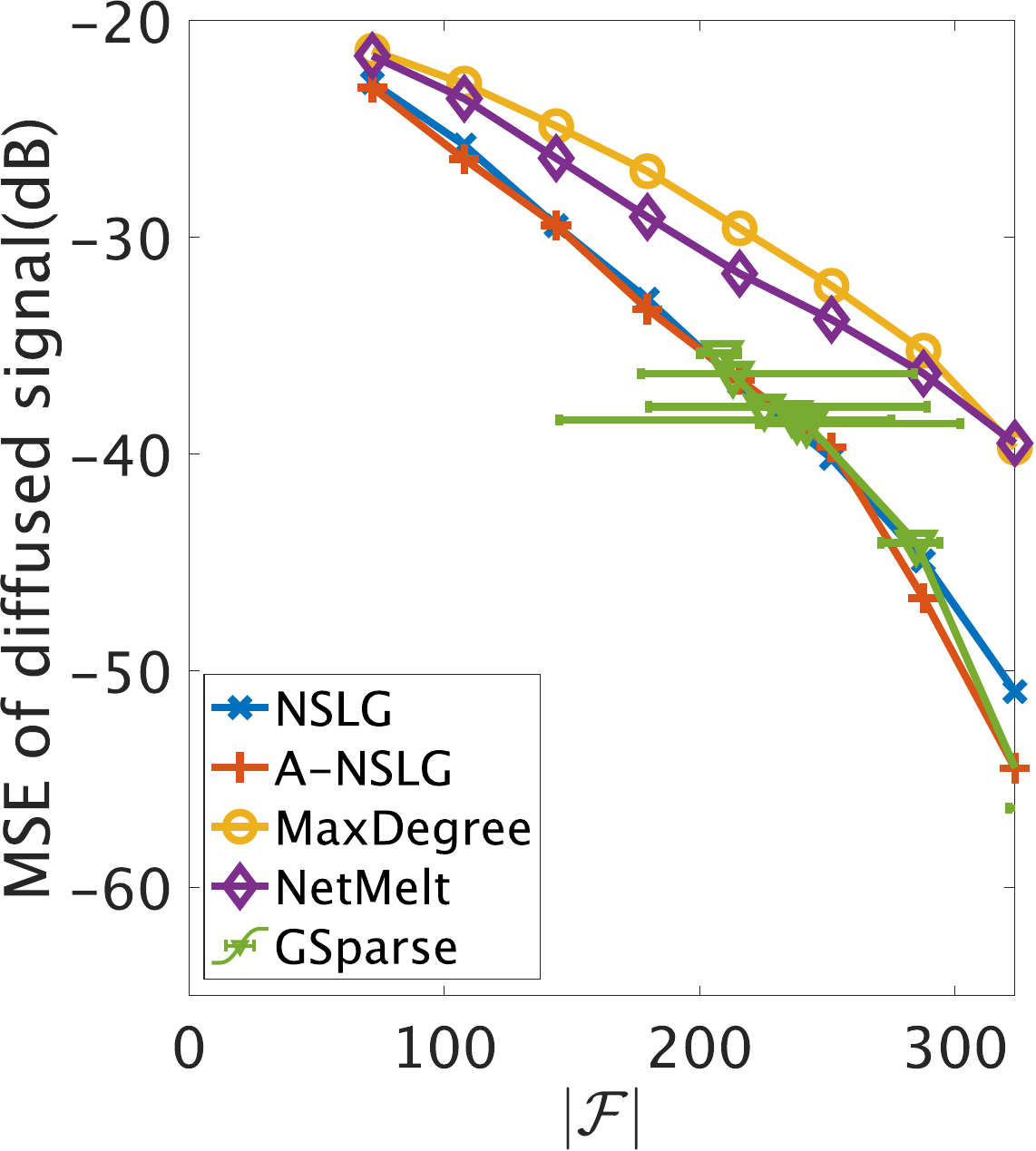}}
    \subfloat[Erd\H{o}s--R\'{e}nyi model]{\includegraphics[width=0.22\linewidth]{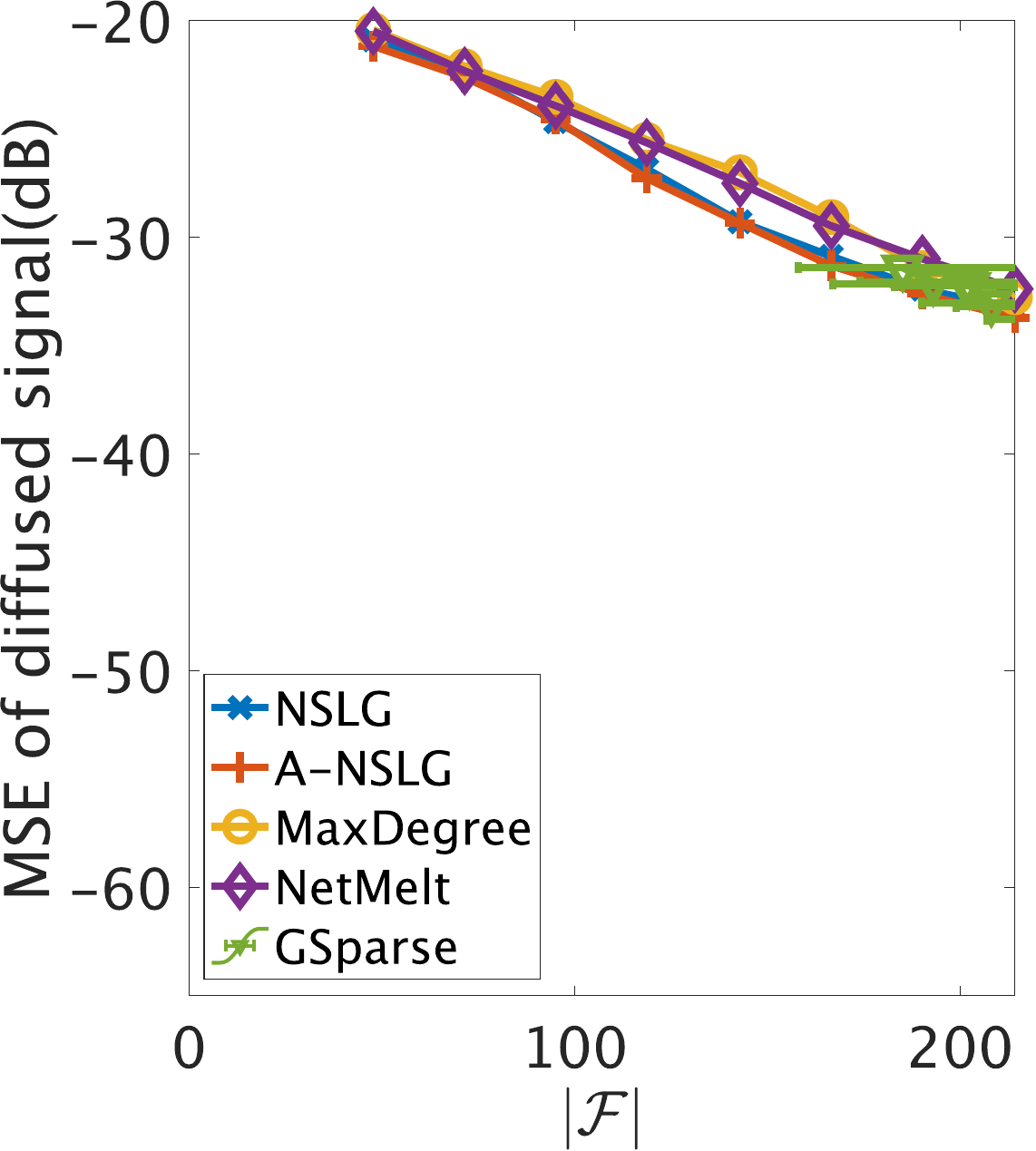}}
    \subfloat[Community graph]{\includegraphics[width=0.22\linewidth]{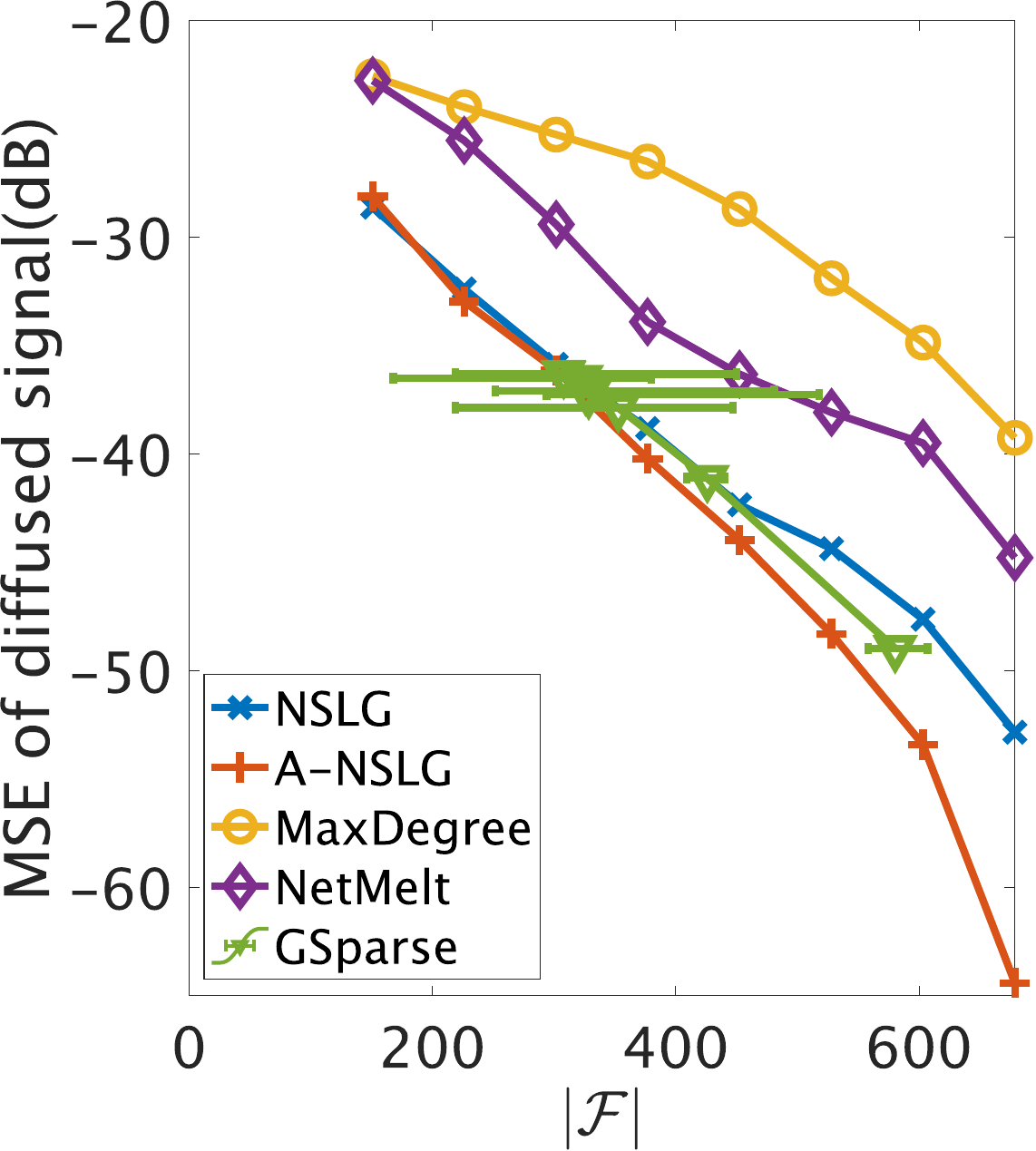}}
    \subfloat[$k$NN graph]{\includegraphics[width=0.22\linewidth]{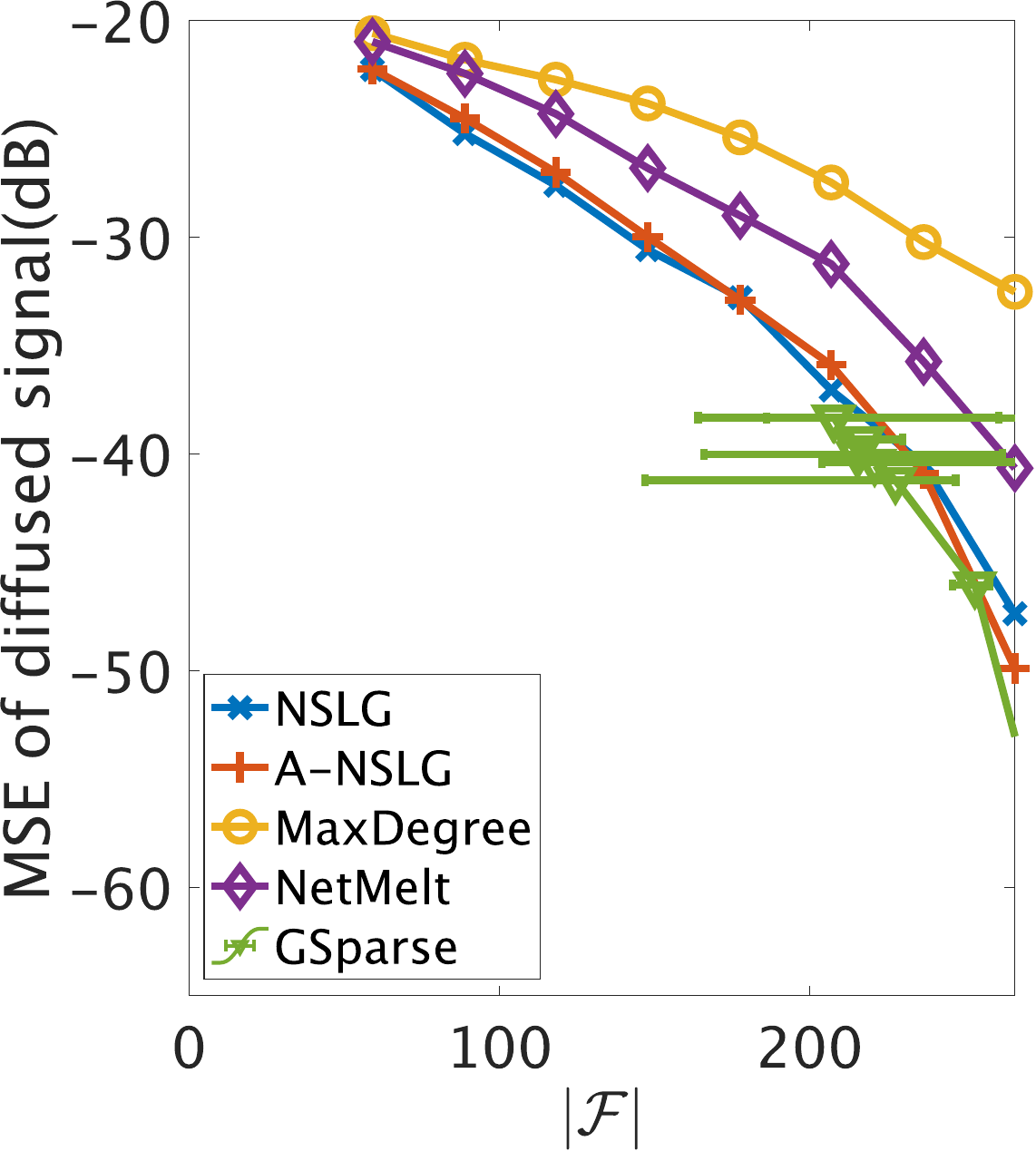}}
    \caption{Comparison of objective performances of sparsified graphs. (a)--(d): Normalized edge weight reconstruction errors. (e)--(h): MSEs of diffused signals in dB. Averaged results after $10$ runs are shown. The horizontal lines of GSparse denote the variations of $|\Fc|$ (i.e., the minimum/maximum number of edges) in the experiment.}
    \label{fig:exp2}
\end{figure*}

\section{Experiments}\label{sec: experiments}
In this section, we perform the proposed edge sampling method for graph sparsification both for synthetic and real-world data to validate the effectiveness of the proposed approach.

\subsection{Synthetic Data}\label{sec: exp-setup}
\subsubsection{Setup}
In this experiment, we use the following weighted graphs with $N  = 100$:
\begin{itemize}
    \item Random sensor network: $|\Ec| = 360$, 
    \item Random graph based on Erd\H{o}s--R\'{e}nyi model with a probability of connection of nodes $0.1$: $|\Ec| = 238$,
    \item Community graph with $5$ communities: $|\Ec| = 754$,
    \item $k$NN graph ($k=6$) constructed from a point cloud with $2$ clusters: $|\Ec| = 296$.
\end{itemize}
Figs. \ref{fig:ex-graph}(a)--(d) show the examples of graphs used in the experiments.
Edge weights in the frequency domain are generated with a bandlimited signal model:
\begin{equation}\label{eqn: edge_weight_exp}
    \hat{\wv} = [\hat{\wv_{K}}^\top, \mathbf{0}_{|\Ec|-K}^\top]^\top + \nv,
\end{equation}
where $\hat{\wv_K}$ is a random vector $\mathbb{R}^{K\times1}$ whose elements are drawn from a normal distribution $\Nc(0, \sqrt{0.2})$, $\nv$ is an i.i.d. additive noise vector with $\Nc(0, 0.1)$.
We set $K$ to $\frac{|\Ec|}{10}$ in all of the experiments with synthetic data.
We then transform the edge weights $\hat{\wv}$ into the edge domain by $\wv = \Vm \hat{\wv}$, where $\Vm$ is the eigenvector matrix of the unweighted line graph.

The accuracy of the sparsification is compared in three measures: 1) Edge weight reconstruction error, 2) MSE of diffused random signals, and 3) Cluster inconsistency of spectral clustering.
\begin{enumerate}
    \item \textbf{Edge weight reconstruction error:} 
    We recover removed edge weights with a graph signal reconstruction method based on the bandlimited assumption \cite{sakiyama2019}.
    The edge weight reconstruction error is given by \eqref{eqn: costfunc}.
    We simply use the MATLAB function \textit{gsp\_graph\_interpolate} in GSP Toolbox \cite{perraudin2016} for $\mathrm{Interp}(\cdot)$ in \eqref{eqn: costfunc}.
    If $\Fc$ in $\Gc_1$ is a good abstraction of $\Ec$ in $\Gc_0$, the recovered edge weights should be close to the original ones.
    \item \textbf{MSE of diffused random signals:} 
    The MSE of diffused random signals are computed as follows.
    Let $\Lm_0$ be the graph Laplacian of $\Gc_0$ and $\Lm_1$ be that of $\Gc_1$.
    The diffused signal on $\Gc_k$ ($k \in \{0,1\}$) is represented as follows:
    \begin{equation}\label{eqn: diffused_signal}
        \yv_k := h(\Lm_k) \xv = \Um_k h(\Lambdam_k) \Um_k^\top \xv,
    \end{equation}
    where $\xv \in \{0, 1\}^N$ is the input signal, $h(\Lm_k)$ is the lowpass graph filter on $\Gc_k$, and $\Lm_k := \Um_k \Lambdam_k \Um_k^\top$.
    We set $h(\lambda) = e^{-t\lambda}$ where $t > 0$ is a parameter.
    Nonzero elements in $\xv$ are randomly chosen, and their number is set to be $20$.
    Finally, the MSE of the diffused signal is calculated by
    \begin{equation}
        \text{MSE}(\yv_0, \yv_1) = \frac{1}{N}\|\yv_0 - \yv_1\|_2^2.
    \end{equation}
    If $\Gc_1$ preserves the original structure, the diffused signal values on $\Gc_1$ will become similar to those on $\Gc_0$.
    This results in a low MSE.

    \item \textbf{Cluster inconsistency of spectral clustering:} 
    Cluster inconsistency $C$ is represented as follows:
    \begin{equation}
        C = 1 - \frac{1}{N}\sum_{i=0}^{N-1} s_i,
    \end{equation}
    where $s_i$ is the indicator element in which $s_i = 1$ when the cluster assigned to node $i$ after edge sampling is the same as that of the original graph and $0$ otherwise.

    We utilize a spectral clustering method \cite{NIPS2001_801272ee} for the experiment.
    If the graph spectrum after edge sampling maintains the original one, the clusters assigned before and after sampling will coincide.
    Therefore, $C$ is expected to be small.
\end{enumerate}

In the proposed edge sampling method, we use FastGSSS \cite{sakiyama2019} as a sampling set selection of the line graph.
The method using FastGSSS in the proposed framework (Section \ref{sec: edge_sampling}) and its faster version (Section \ref{sec: acceleration}) are abbreviated as \textit{NSLG} (Node Sampling on Line Graph) and \textit{A-NSLG} (Accelerated Node Sampling on Line Graph), respectively.
For the proposed methods, we set the approximation degree of the CPA to $6$.

The performance is compared with the following edge sampling methods:
\begin{itemize}
    \item \textit{MaxDegree} (deterministic): Greedy selection. Edges having the largest $k_m + k_n$ are selected one by one.
    \item \textit{NetMelt} (deterministic) \cite{chen2016}: Edge selection based on the score calculated from the eigenvectors of $\Lm_0$.
    \item \textit{GSparse} (random) \cite{spielman2011}: Graph sparsification based on effective resistances, where a random selection of edges is performed based on a probability proportional to the effective resistance of $\Gc_0$.
\end{itemize}
The proposed methods and the first two alternative methods are deterministic approaches: The number of edges is specified before edge sampling, and the edges to be removed are fixed as long as the graph is fixed.
In contrast, \cite{spielman2011} is a random approach that requires a sparsification parameter between $\frac{1}{\sqrt{N}}$ and 1.
In other words, the random method cannot determine the number of removed edges.
Note that, even under the same parameter, the removed edge positions of \textit{GSparse} are changed in each realization, and their number can vary due to random selection and replacement.

\begin{figure}[t]
    \centering
    \subfloat[Community graph]{\includegraphics[width=0.44\linewidth]{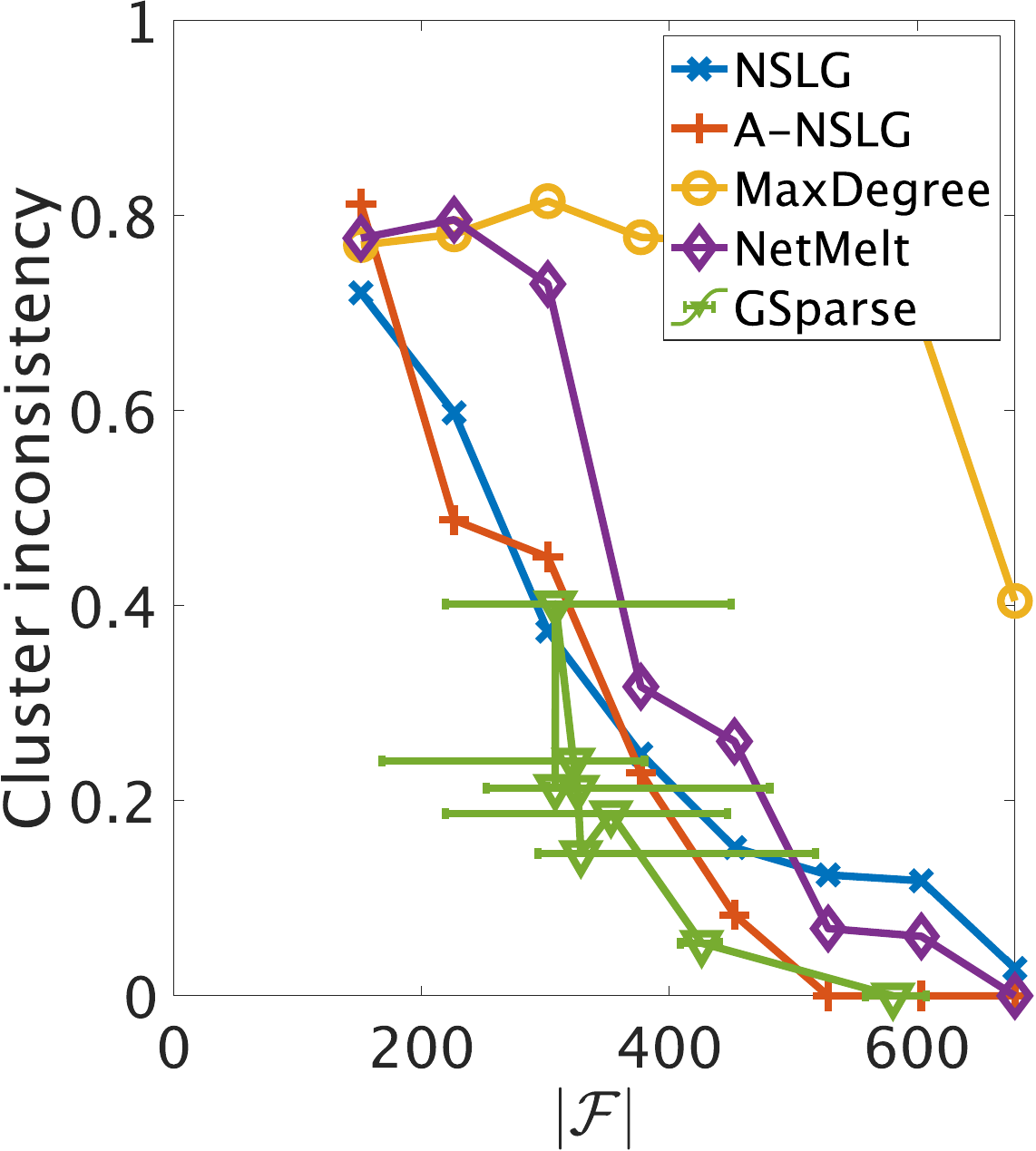}}
    \subfloat[$k$NN graph]{\includegraphics[width=0.44\linewidth]{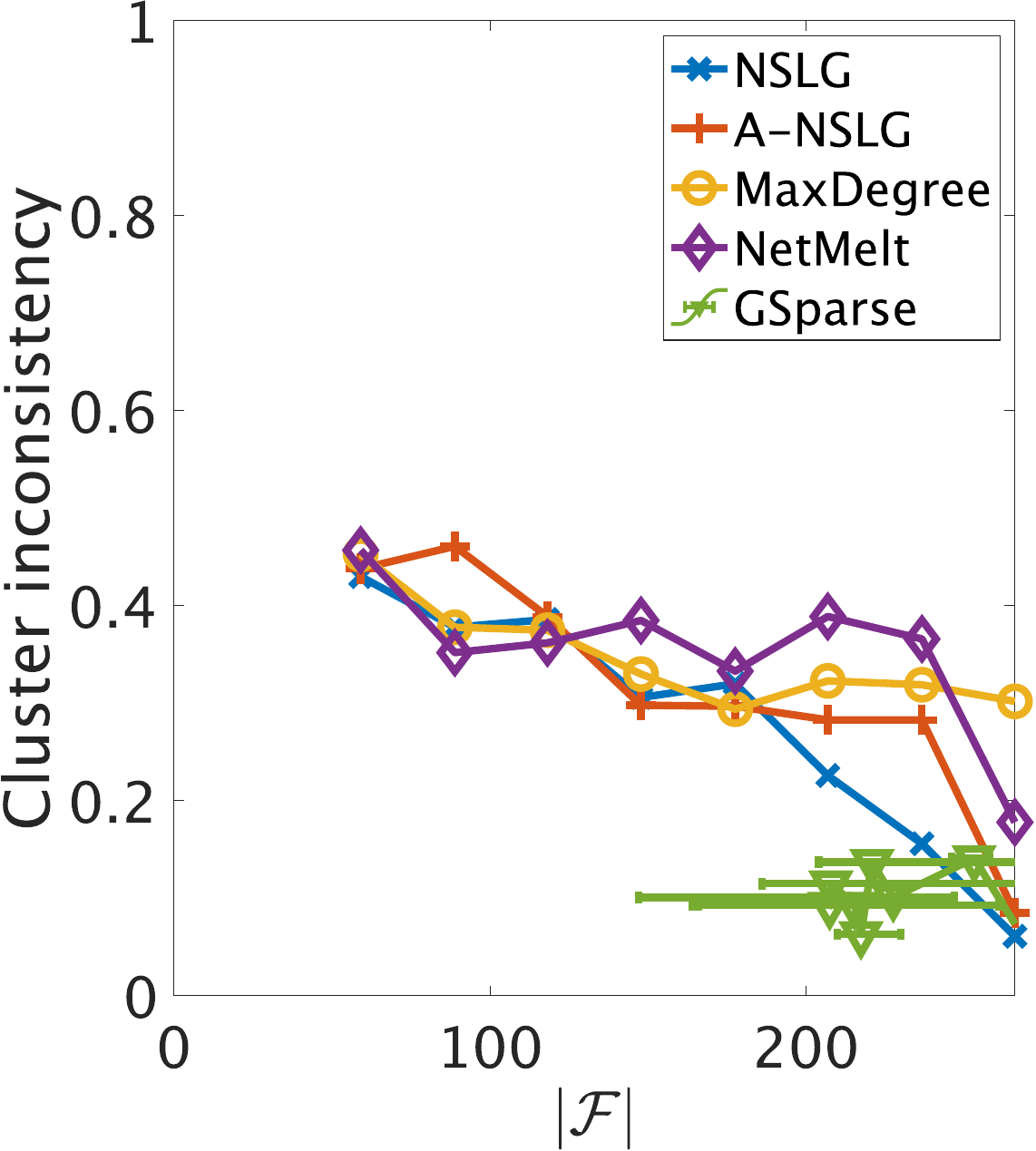}}
    \caption{Comparison of cluster inconsistency of sparsified graphs. Averaged results after $10$ runs are shown. The horizontal lines of GSparse denote the variations of $|\Fc|$ (i.e., the minimum/maximum number of edges) in the experiment.}
    \label{fig:exp3}
\end{figure}

\begin{figure*}[t]
\centering
\subfloat[Original graph-$\xv$]{\includegraphics[width=0.24\linewidth]{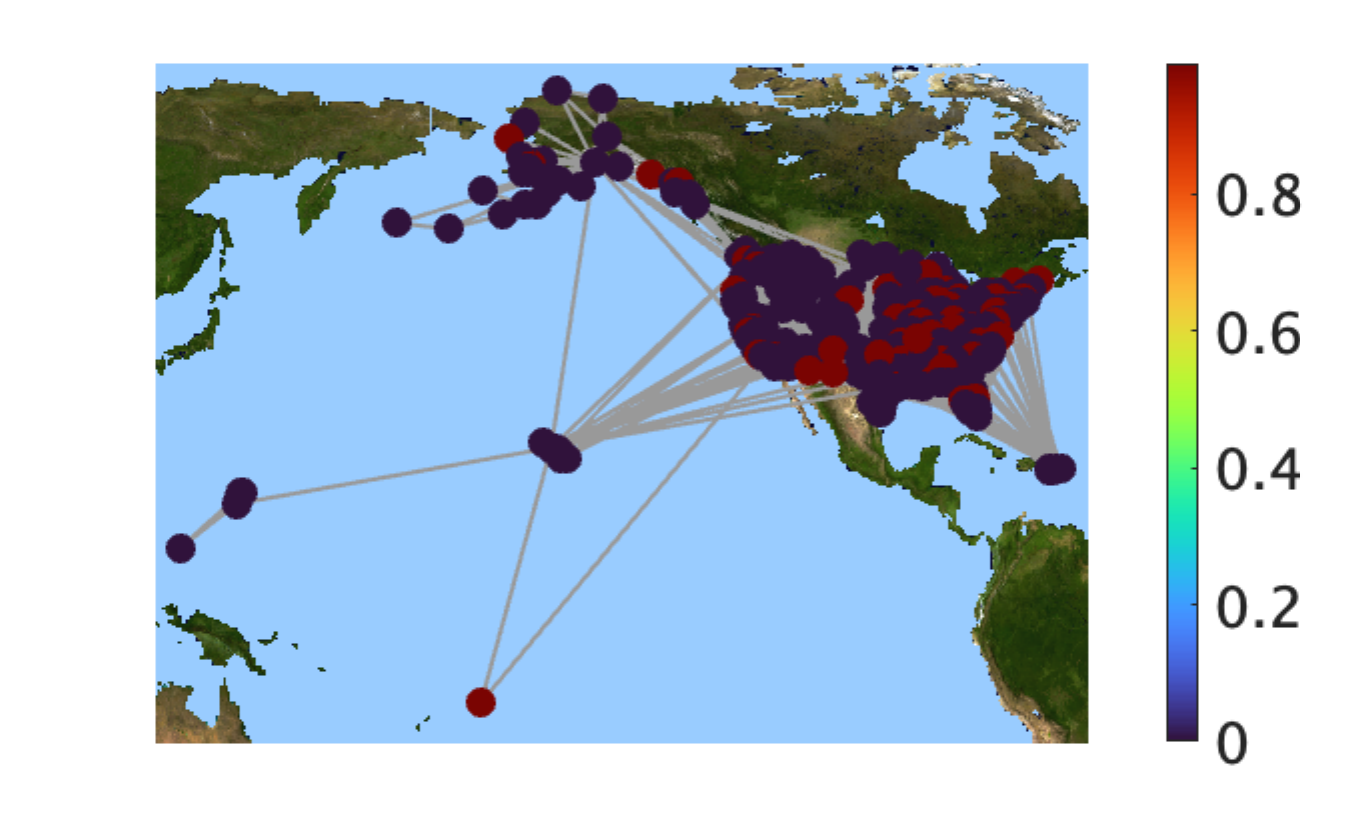}}
\subfloat[Original graph-$\yv_0$]{\includegraphics[width=0.24\linewidth]{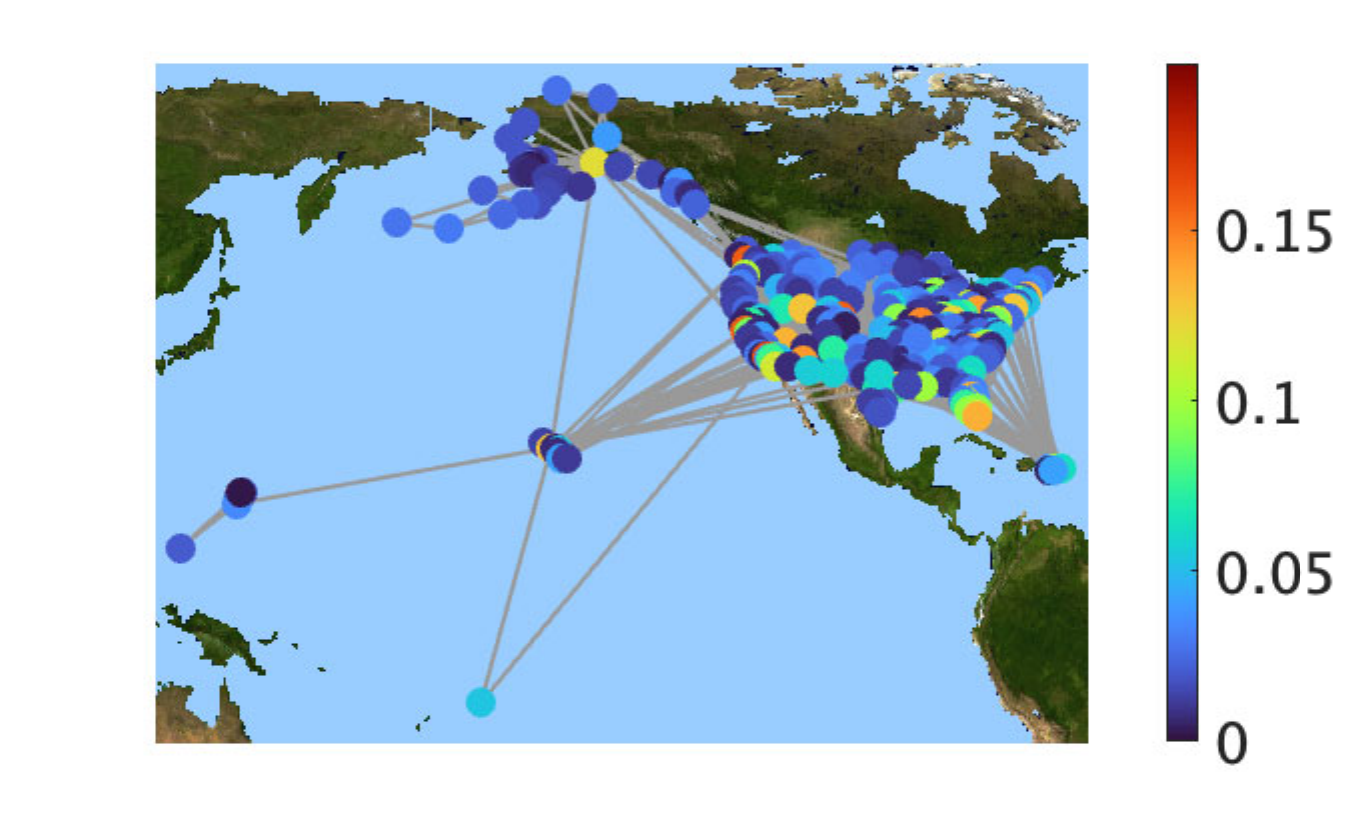}}
\subfloat[NSLG-$\yv_1$]{\includegraphics[width=0.24\linewidth]{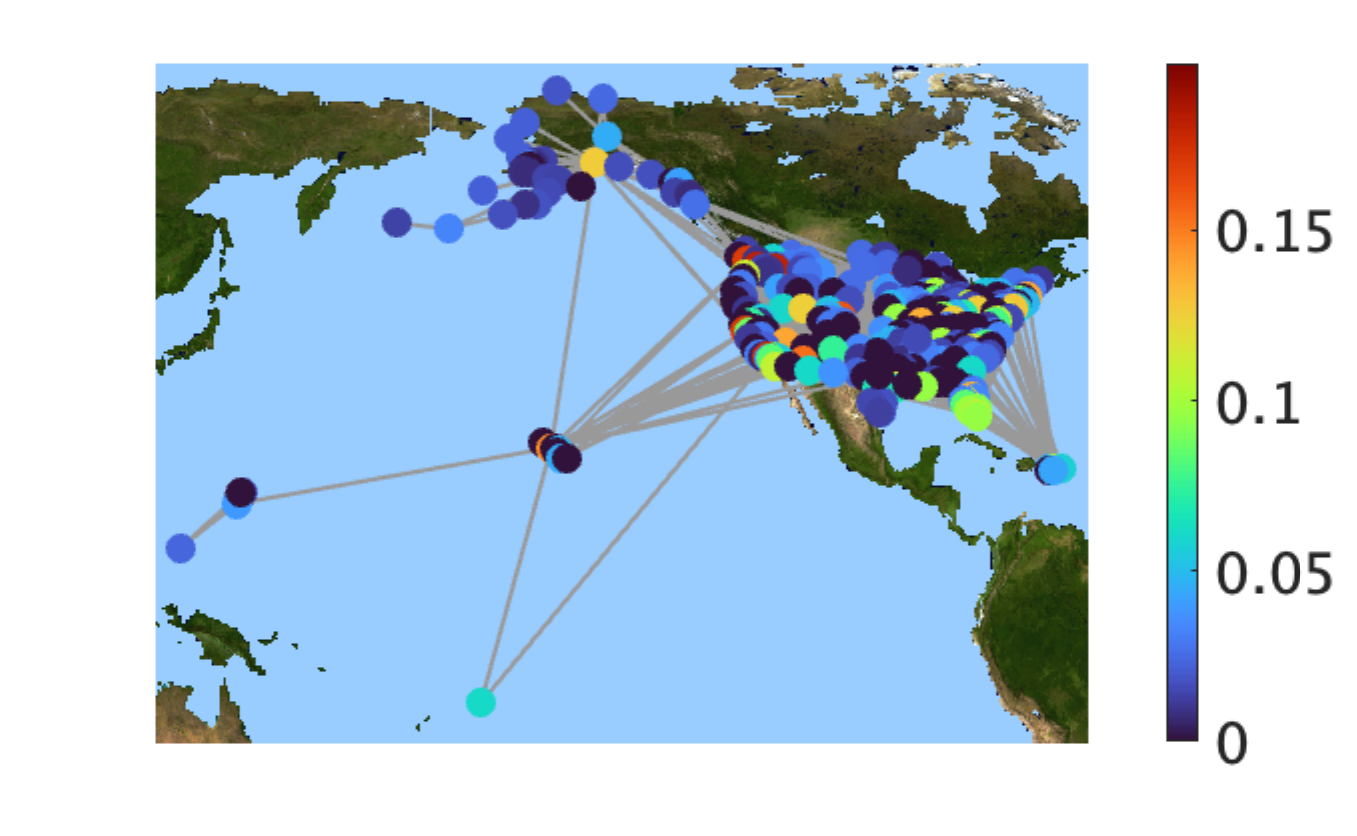}}
\subfloat[Proposed-Faster-$\yv_1$]{\includegraphics[width=0.24\linewidth]{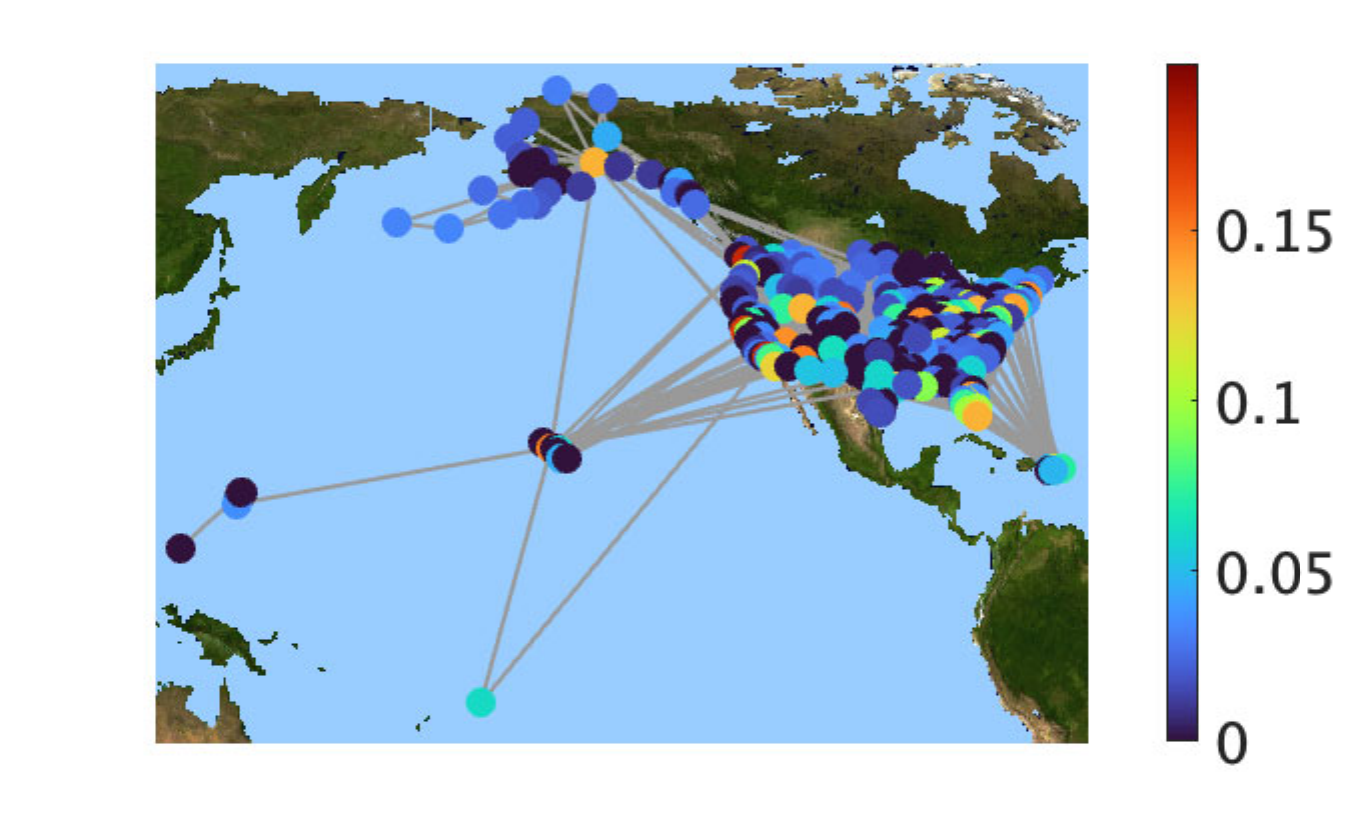}}\\
\subfloat[MaxDegree-$\yv_1$]{\includegraphics[width=0.24\linewidth]{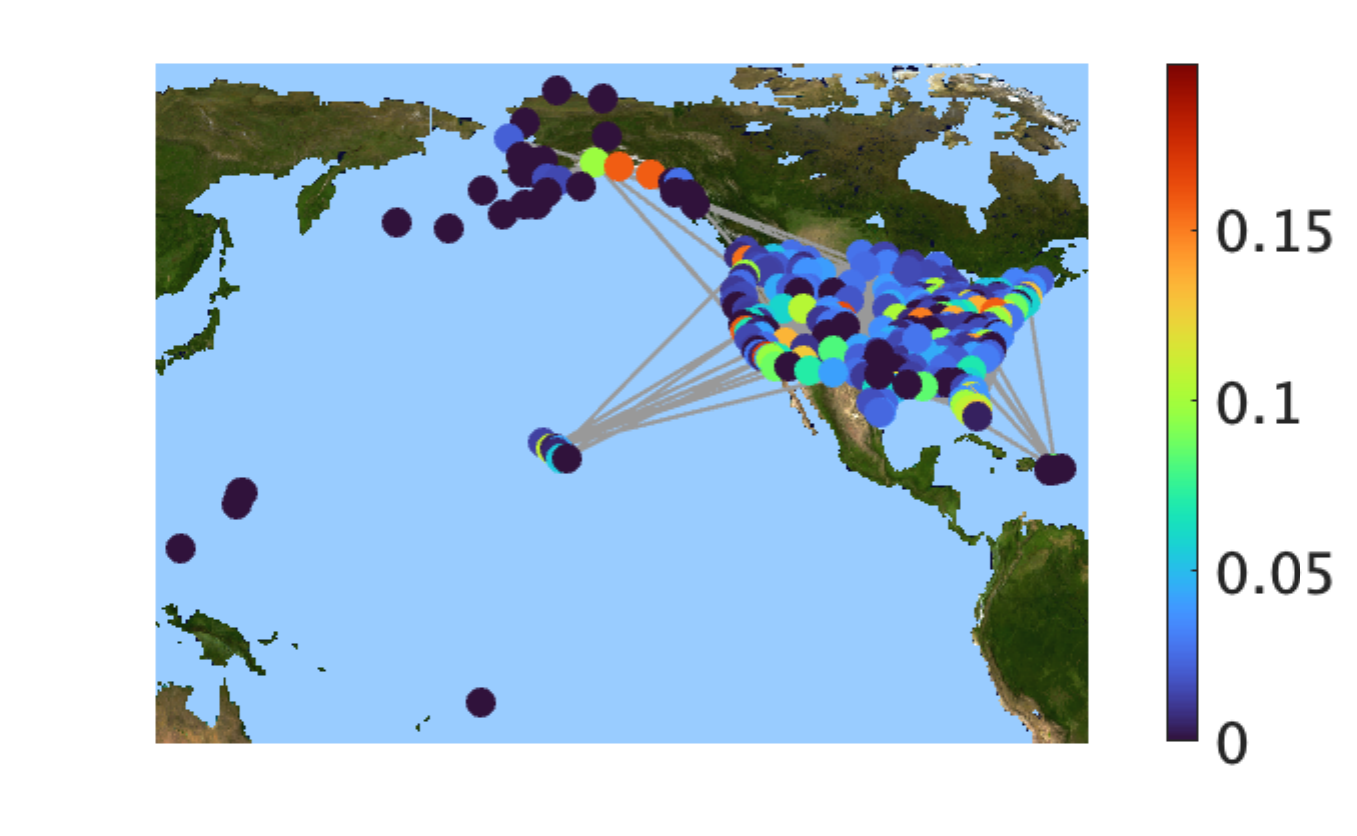}}
\subfloat[NetMelt-$\yv_1$]{\includegraphics[width=0.24\linewidth]{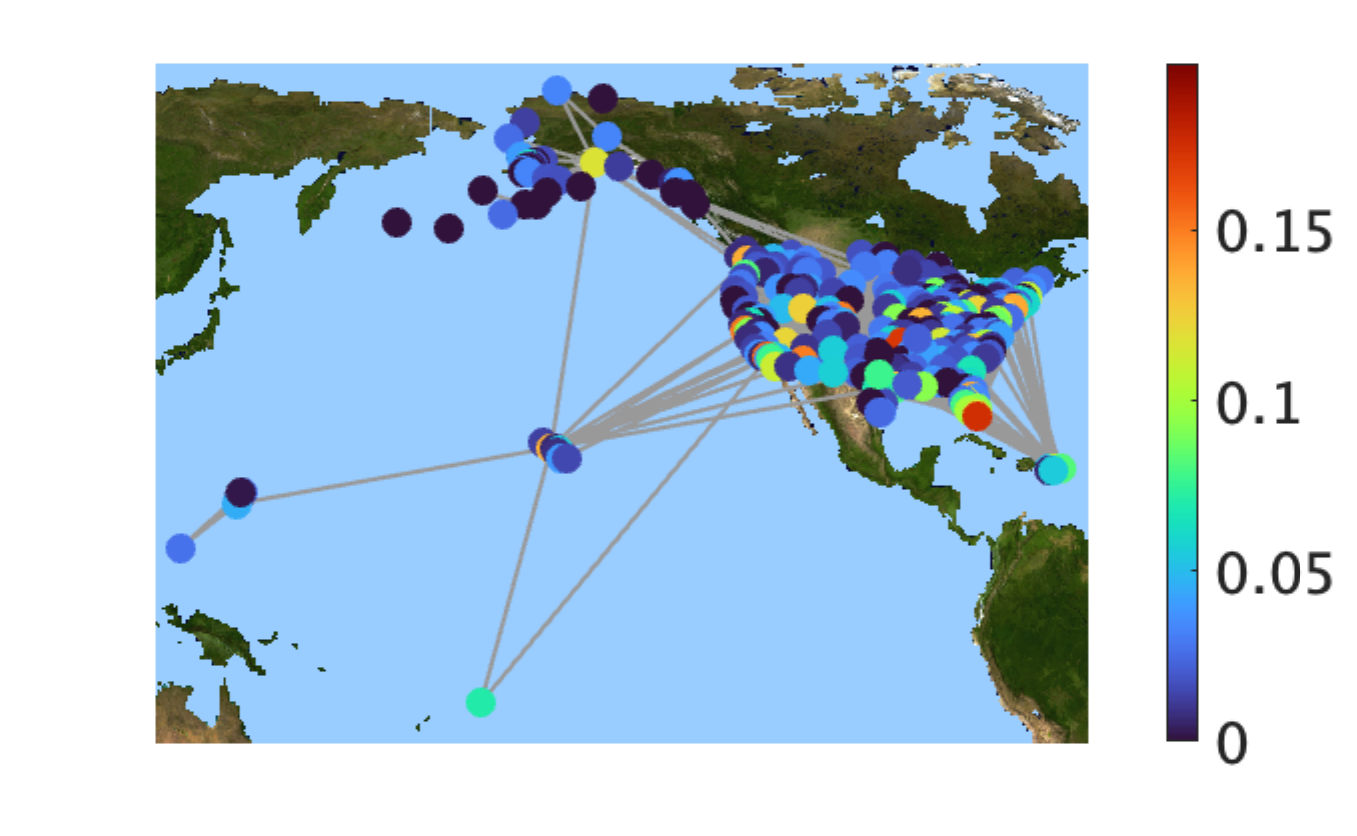}}
\subfloat[GSparse-$\yv_1$]{\includegraphics[width=0.24\linewidth]{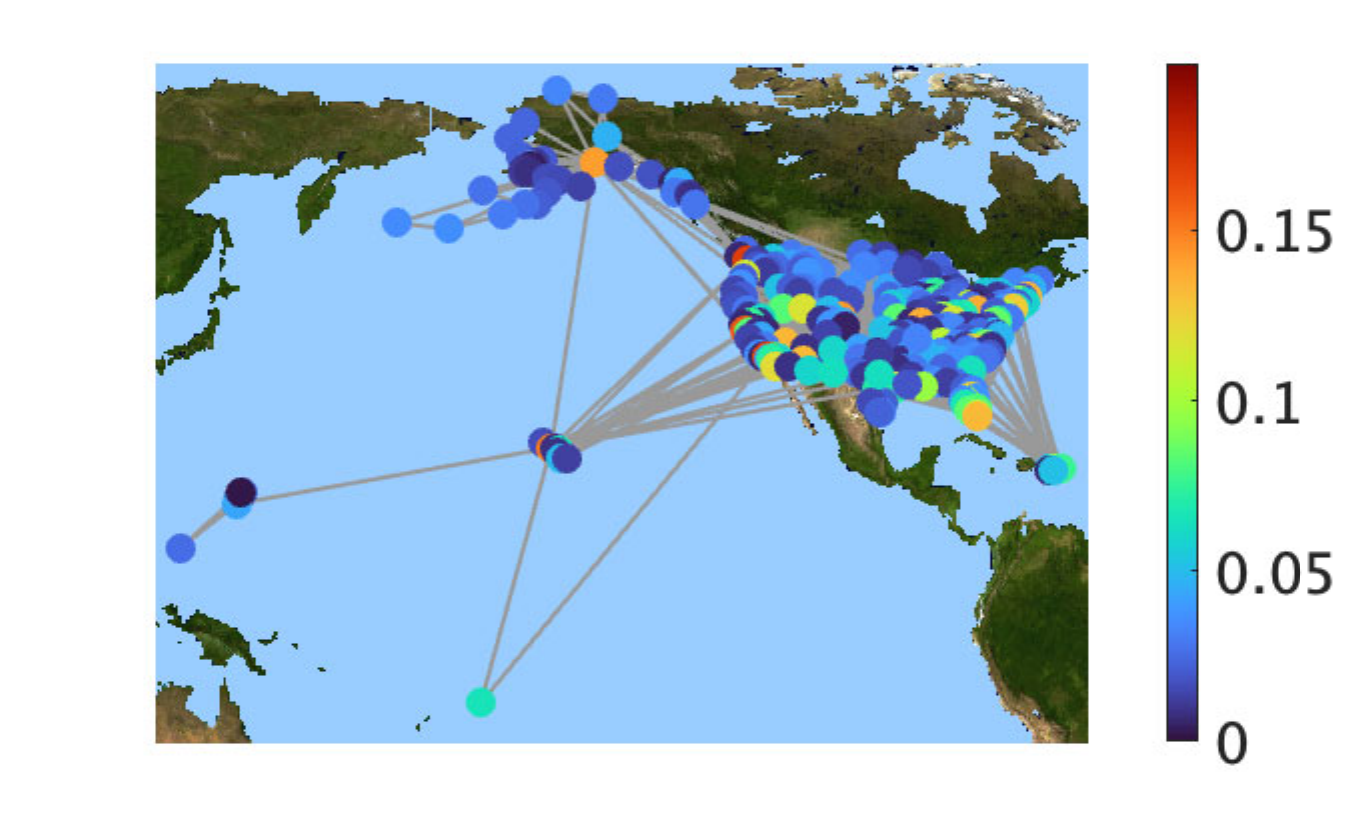}}
\caption{Graph sparsification and diffusion example: USAir97 dataset. Diffused signals are also shown.}
\label{fig:exp1-USAir97}
\end{figure*}

\subsubsection{Experimental Results}
Fig. \ref{fig:exp1-community} shows the sparsified graphs by sampling the edges in half, as well as the diffused signals on them.
The proposed methods and GSparse are almost connected, while MaxDegree and NetMelt often isolate nodes.
It is also observed that the diffused signals of Figs. \ref{fig:exp1-community}(b)--(d), and (g) are similar to each other.

Figs. \ref{fig:exp2}(a)--(d) show $f(\Gc_1)$ in \eqref{eqn: costfunc} as functions of $|\Fc|$.
As previously mentioned, the number of removed edges by GSparse varies even under the same parameter.
Therefore, we also illustrate such a variation in the figure.
The proposed methods show consistently lower edge weight reconstruction errors than the other methods.

Figs. \ref{fig:exp2}(e)--(h) show the average MSEs of the diffused signal according to $|\Fc|$ in the sparsified graph.
As observed in the sparsified graphs, the proposed methods and GSparse present comparable MSEs and are better than MaxDegree and NetMelt.
The proposed methods enable a wide range of edge sparsification factors because they can specify the number of edges thanks to deterministic sampling.
In contrast, GSparse has a small admissible range of $|\Ec|$.
As previously mentioned, the number of removed edges by GSparse significantly varies under the same parameter, where its range sometimes exceeds $200$, which is about one-third of $|\Ec|$.

Fig. \ref{fig:exp3} compares the cluster inconsistency $C$ for each number of sampled edges $|\Fc|$.
The proposed methods have a lower inconsistency rate due to sampling than the other alternative methods.
In other words, the proposed methods select the important edges in the original graph and reduce the change of the graph spectrum due to sparsification.
The results by NSLG and the A-NSLG oscillate due to the use of the same parameters regardless of the number of sampled edges $|\Fc|$.

\subsection{Real Data}
We also perform sparsification experiments through edge sampling with real data.

\subsubsection{Setup}
We use the USAir97 dataset\cite{nr} as the actual network.
USAir97 is a dataset whose nodes and edges represent airports and the flights between them.
The number of nodes and edges are $332$ and $2162$, respectively.
The graph is an undirected weighted graph, where the edge weights represent the number of flights between airports.

In this experiment, we set $K$ in \eqref{eqn: edge_weight_exp} and the number of nonzero elements in the input signal $\xv$ in \eqref{eqn: diffused_signal} to $\frac{|\Ec|}{60}$ and $\frac{|\Ec|}{5}$, respectively.
The other parameters were set to the same values as in the experiments with synthetic data.
We compare the performance of the proposed method with the same method on synthetic data experiments.
As evaluation criteria, we use the edge weight reconstruction error and the MSE of the diffused random signal (see Section \ref{sec: exp-setup}).

\subsubsection{Experimental Results}
\begin{figure}[t]
    \centering
    \subfloat[Edge weight reconstruction error]{\includegraphics[width=0.44\linewidth]{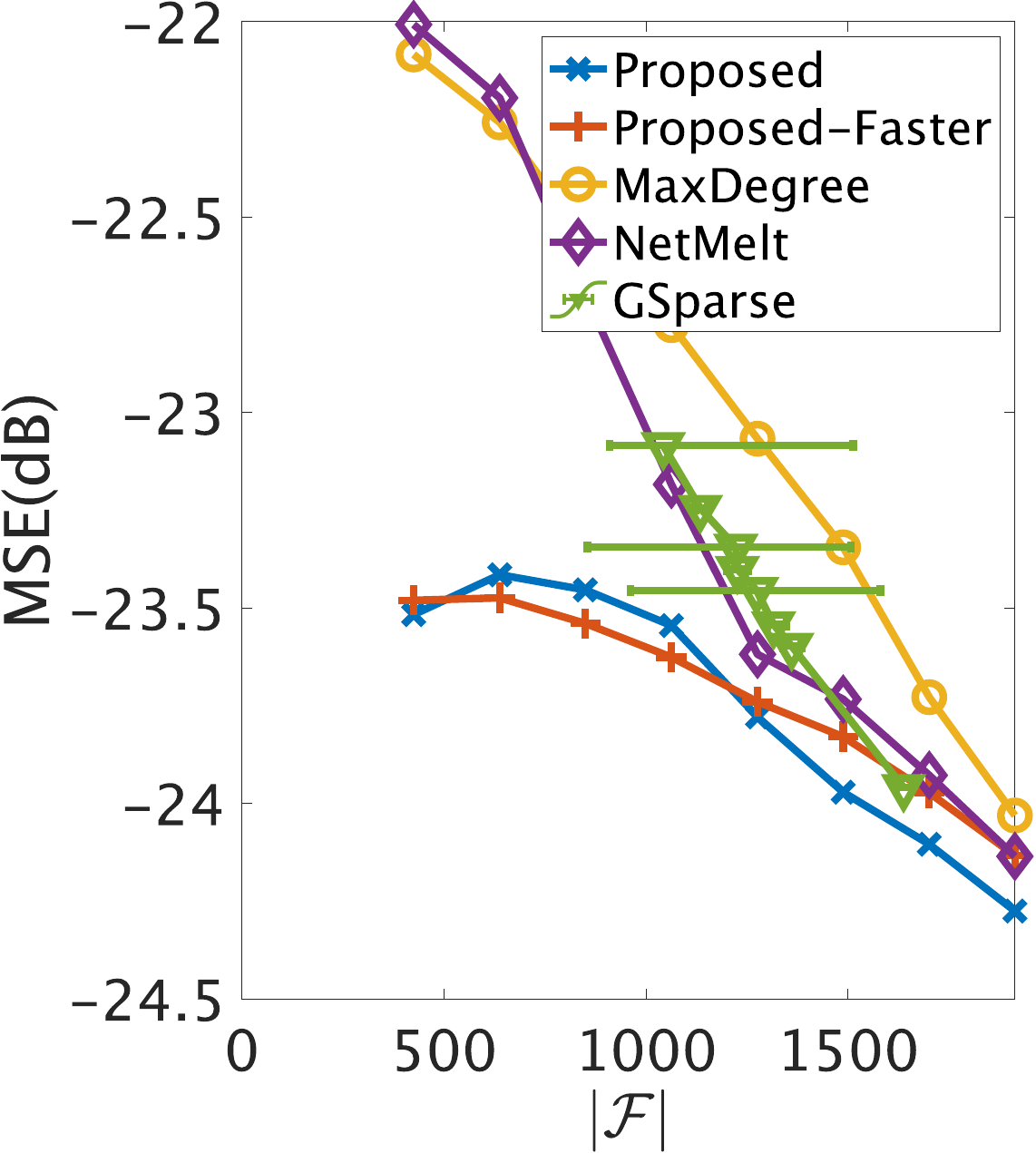}}
    \subfloat[MSE of diffused signals]{\includegraphics[width=0.44\linewidth]{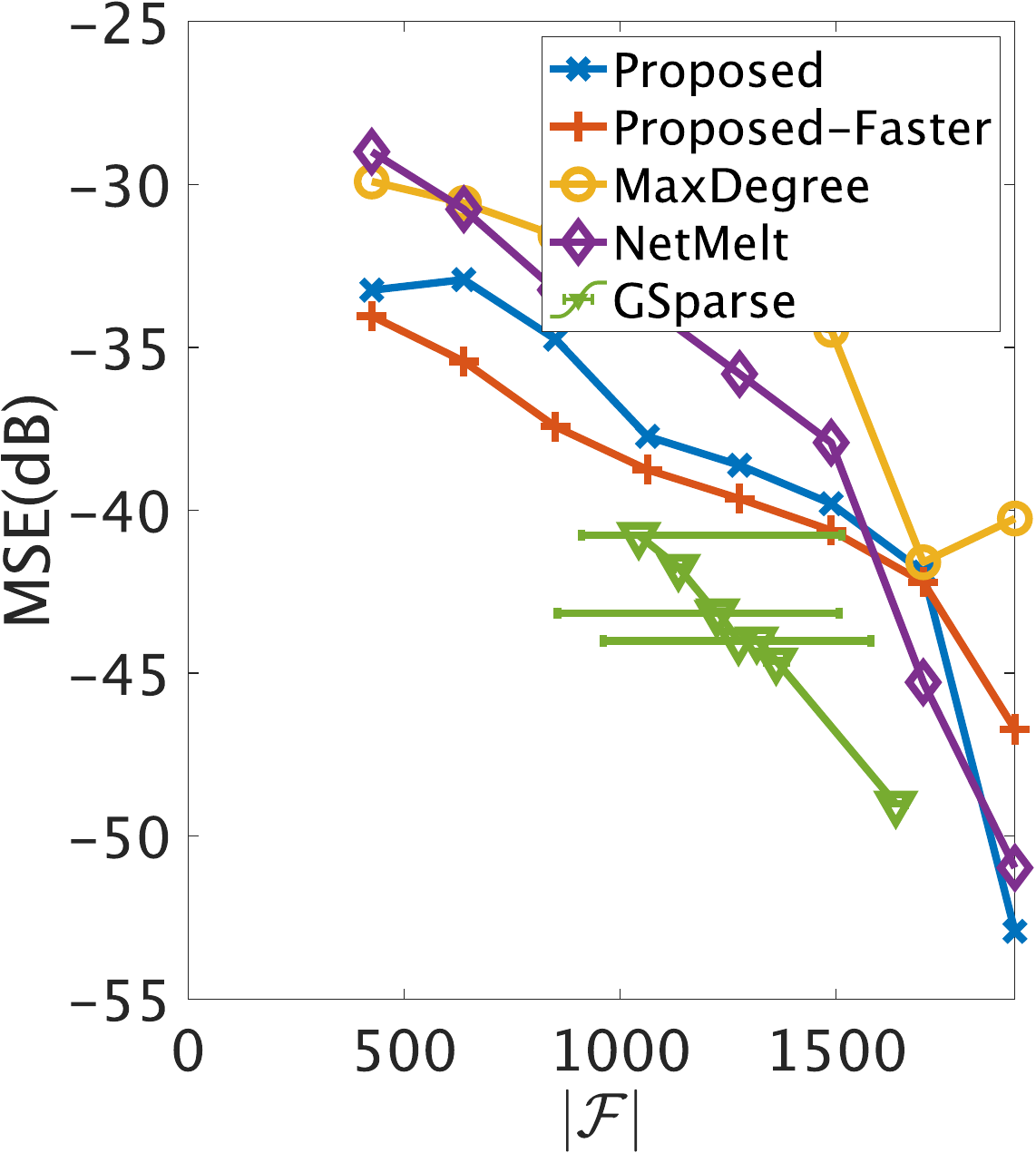}}
    \caption{Comparison of objective performances of sparsified graphs. (a): Normalized edge weight reconstruction errors. (b): MSE of diffused signals in dB. Averaged results after $10$ runs are shown. The horizontal lines of GSparse denote the variations of $|\Fc|$ (i.e., the minimum/maximum number of edges) in the experiment.}
    \label{fig:exp2-USAir97}
\end{figure}

Fig. \ref{fig:exp1-USAir97} shows the visualization of graph sparsification results and diffused signals on these sparsified graphs.
As in the experiments on synthetic data, both of the proposed methods can remove the half number of edges (i.e., $1081$ edges) without isolated nodes.
Fig. \ref{fig:exp2-USAir97}(a) and (b) show the 
edge weight reconstruction error and the MSE of the diffused random signals, respectively.
Overall, our proposed methods present satisfactory results in MSEs and flexibility on the number of edges.

\section{Conclusion} \label{sec: conclusion}
In this paper, we propose an edge sampling method based on graph sampling theory.
We first convert the original graph into a line graph and perform a sampling set selection of graphs.
The edge smoothness characteristic is converted to signal smoothness via graph conversion.
In this paper, we also propose a further acceleration method for our edge sampling approach by using a spectral relationship between a graph Laplacian of the line graph and an edge Laplacian.
The experimental results on edge sparsification reveal the effectiveness of the proposed method against some alternative approaches.

\bibliographystyle{IEEEtran}
\bibliography{IEEEabrv, refs}

\end{document}